\documentclass[lettersize,journal]{IEEEtran}
\usepackage{enumitem}
\usepackage{wrapfig}
\usepackage{amsmath,amsfonts}
\usepackage[caption=false,font=normalsize,labelfont=sf,textfont=sf]{subfig}
\usepackage{stfloats}
\usepackage{graphicx}
\usepackage{cite}
\usepackage{url}
\usepackage{xcolor}

\begin{document}

\title{Interlocking Tiling of 3D Finite Domains}
\author{Xuyang Ma, Rulin Chen, Chaewoon Kim, Ying He, Peng Song

\thanks{
Xuyang Ma and Rulin Chen are co-first authors. 
Corresponding author: Peng Song.

Xuyang Ma, Chaewoon Kim and Peng Song are with the Singapore University of Technology and Design, Singapore 487372 (e-mail: xuyang\_ma@mymail.sutd.edu.sg; chaewoon\_kim@mymail.sutd.edu.sg; peng\_song@sutd.edu.sg).

Rulin Chen is with the Beijing Normal-Hong Kong Baptist University, China 519088 (e-mail: rulinchen@bnbu.edu.cn).

Ying He is with the Nanyang Technological University, Singapore 639798 (e-mail: yhe@ntu.edu.sg).

}
}

\maketitle

\begin{abstract}
A $K$-hedral tiling of a 3D finite domain is a covering of the domain with tiles without gaps or overlaps, where each tile is congruent to one of the $K$ distinct shapes called \emph{prototiles}.
To use 3D $K$-hedral tilings in practice, they need to satisfy two requirements.
First, $K$ should be small to reduce the fabrication cost (e.g., through reusable molds) and simplify the (dis)assembly process.
Second, a $K$-hedral tiling result should be \emph{structurally stable}, in particular via interlocking, where all the tiles are immobilized by their geometric arrangement, except the key.
Since these two requirements are often in conflict, existing approaches always trade one for the other.

In this work, we formulate a new tiling problem called \emph{interlocking tiling}, aiming to generate 3D $K$-hedral tilings that satisfy the above two requirements simultaneously.
Given a 3D finite domain represented as a voxel grid, our goal is to compute a small prototile set whose instances exactly tile the input domain while ensuring that the tiling result forms an interlocking assembly.
To solve this challenging problem, we propose a two-level computational approach that hierarchically decomposes the input domain into interlocking parts (domain-level partitioning) and further partitions each part into interlocking tiles (part-level partitioning).
To guide the part-level partitioning, we propose a new graph representation with regular structures to facilitate the generation of tiles with identical shapes while ensuring that they are interlocking.
Our approach allows generating interlocking tilings that exactly cover 3D finite domains with diverse shapes while achieving high prototile reuse.
We compared our approach with state-of-the-art methods to demonstrate its superiority, and validated the practicality of our interlocking tiling results via fabricating two of them.
\end{abstract}

\begin{IEEEkeywords}
$K$-hedral tiling, interlocking assembly, geometry modeling, structural stability
\end{IEEEkeywords}

\begin{figure*}[!t]
	\centering
	\includegraphics[width=0.99\textwidth]{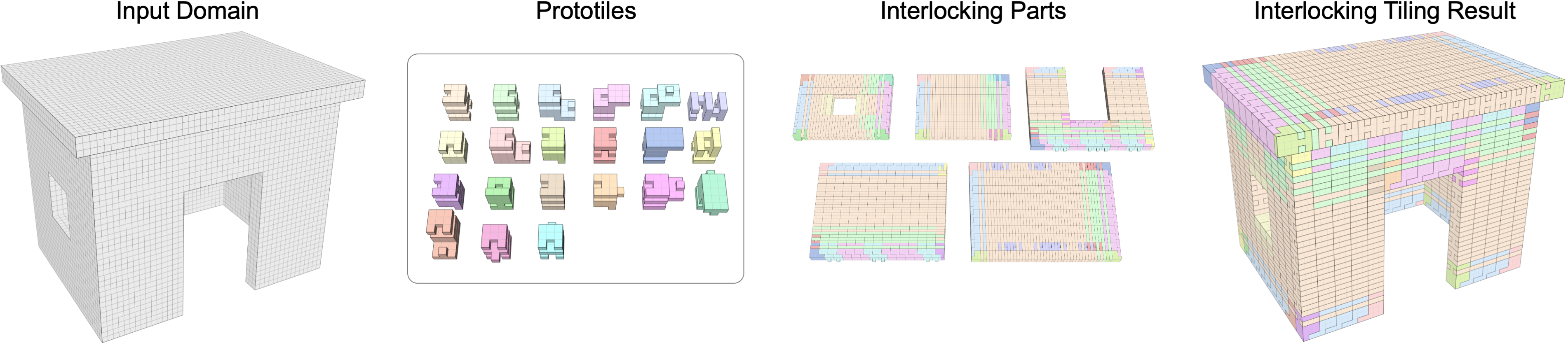}
	\caption{
		Given (left) a 3D finite domain, our approach computes (middle left) 21 prototiles whose instances (i.e., 671 tiles) exactly tile the domain with two-level interlocking.
		Specifically, the tiles are first assembled into (middle right) a few subassemblies (i.e., parts), where each of them is interlocking.
		These parts are then assembled into (right) an interlocking tiling result that exactly covers the input domain without gaps or overlaps.
	}
	\vspace*{1.0mm}
	\label{fig:teaser}
\end{figure*}

\section{Introduction}
\label{sec:intro}

A tiling is a covering of a domain, without gaps or overlaps, using a countable set of geometric shapes called {\em tiles}.
A $K$-hedral tiling is a tiling in which every tile is congruent to one of $K$ distinct shapes called {\em prototiles}~\cite{Kaplan-2009-TilingTheory}.
$K$-hedral tilings of 3D finite domains have been used in a variety of real-world applications, including furniture, architecture, and recreation.
In these applications, it is highly desirable to have a small $K$.
This is because a small prototile set can substantially reduce the fabrication cost (e.g., through reusable molds), simplify the (dis)assembly process, ease maintenance and replacement, and make the tiling aesthetically pleasant with more
congruent appearance.

\begin{figure}[!t]
	\includegraphics[width=8.35cm]{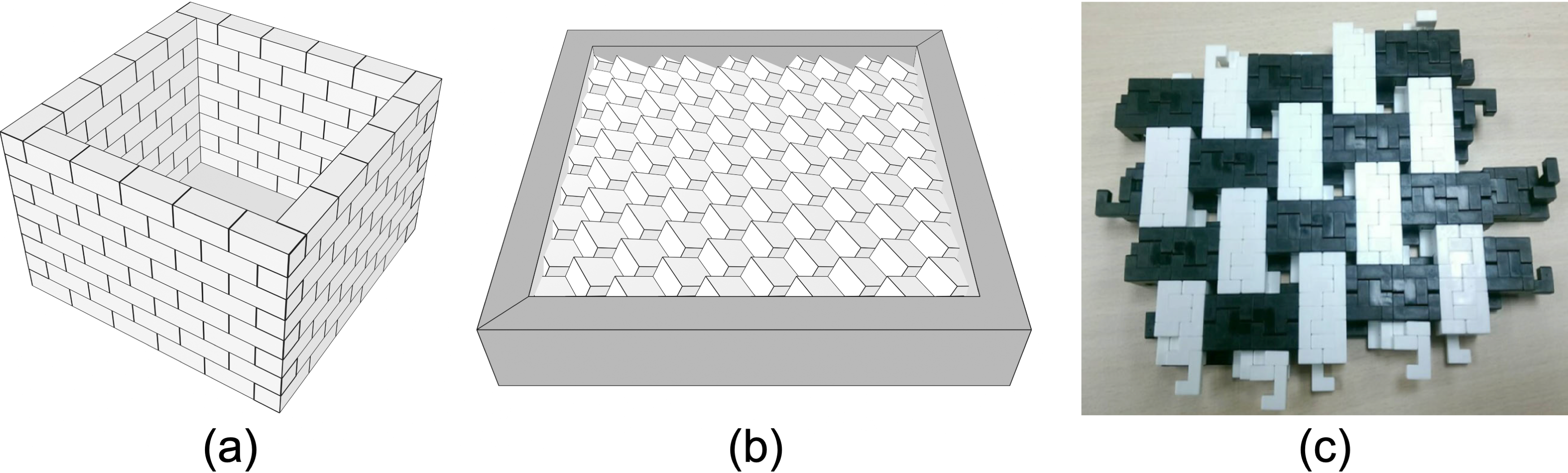}
	\caption{Examples of $K$-hedral tilings of 3D finite domains.
		(a) Masonry walls where bricks are connected by mortar.
		(b) A planar topological interlocking structure with cubes~\cite{Dyskin-2019-TIMaterials}.
		(c) An interlocking structure built with predefined SL blocks~\cite{Shih-2016-SLBlocks}.}
	\label{fig:3D_tiling_examples}
\end{figure}

In addition to having a small $K$, the other challenge is to ensure that $K$-hedral tilings of 3D finite domains are {\em structurally stable} such that they can be used in the real world.
Despite its importance, the problem of modeling structurally stable $K$-hedral tilings of 3D finite domains remains largely underexplored.
For example, masonry walls built with bricks are monohedral tilings with cuboids; see Figure~\ref{fig:3D_tiling_examples}(a).
To ensure structural stability, the bricks are connected via mortar, which however discourages disassembly and maintenance of the structure.
To overcome this challenge, some researchers proposed to use a large frame to fix the boundary of a planar structure of tiles (e.g., cubes), a concept called topological interlocking~\cite{Dyskin-2019-TIMaterials}; see Figure~\ref{fig:3D_tiling_examples}(b).
Yet, the approach cannot be generalized to modeling $K$-hedral tilings of spatial domains. %

Given a 3D finite domain, we aim to model structurally stable $K$-hedral tilings of the domain using a small prototile set (i.e., a small $K$).
To make the problem tractable, we assume that the input domain is represented as a voxel grid.
The major challenge of this problem arises from two conflicting goals.
On the one hand, we want to have a small number of prototiles whose instances (i.e., tiles) can exactly tile the input domain, aiming to maximize reusability of the prototiles.
On the other hand, the tiling result has to be structurally stable via interlocking among the tiles.
Existing approaches cannot achieve these two goals simultaneously and thus trade one for the other.
They either prescribe a small set of prototiles and produce only an approximation of the input domain~\cite{Zhang-2016-InterlockVoxel,Shih-2016-SLBlocks},
or partition the input domain into interlocking tiles that are almost all distinct in shape (i.e., no prototile reuse)~\cite{Song-2012-InterCubes,Wang-2018-DESIA,Chen-2022-HighLevelPuzzle}.

To address the above challenge, we make the following technical contributions:
\begin{itemize}[leftmargin=*]
	\item
	In Section~\ref{sec:problem}, we formulate a new tiling problem, called {\em interlocking tiling of 3D finite domains}, which aims to simultaneously (i) minimize the number of prototiles and (ii) ensure that the prototiles' instances are interlocking.

	\vspace*{1.0mm}
	\item
	In Section~\ref{sec:approach}, we propose a hierarchical approach to solve the interlocking tiling problem by decomposing the input domain in two stages: (i) domain-level partitioning, which partitions the input domain into interlocking parts, and (ii) part-level partitioning, which further partitions each part into interlocking tiles.

	\vspace*{1.0mm}
	\item
	In Section~\ref{sec:graph}, we propose Interlocking Tiling Directional Blocking Graphs (IT-DBGs) that encode blocking relations in typical interlocking tilings.
	In Section~\ref{sec:partition}, we then leverage IT-DBGs to guide the partitioning of each part into interlocking tiles while minimizing the number of prototiles.
\end{itemize}

Our approach allows generating interlocking tilings for 3D finite domains with various shapes.
Figure~\ref{fig:teaser} shows an interlocking tiling \emph{House} computed by our approach, which has 671 tiles and 21 prototiles.

\section{Related Work}
\label{sec:related}

\noindent
\textbf{Tiling of finite domains.} \
Recently, researchers are interested in the problem of tiling a 2D finite domain using a set of 2D shapes.
Garvie and Burkardt~\cite{Garvie-2020-TilingFiniteRegions, Garvie-2022-TilingFiniteRegions} implemented forward approaches for tiling a 2D finite domain by using a set of predefined polyominoes.
In contrast, Chen et al.~\cite{Chen-2025-InverseTiling} proposed an inverse tiling approach that tiles a 2D finite domain without prescribing prototile shapes, aiming to minimize the number of prototiles required to tile the domain.
However, all these approaches are tailored to 2D finite domains and cannot directly be extended to 3D finite domains, where the tiling results have to be structurally stable for practical usage.
To address this challenge, we introduce a new concept called {\em interlocking tiling} for modeling structurally stable $K$-hedral tilings of 3D finite domains with a small prototile set.

\vspace*{1.0mm}
\noindent
\textbf{Space-filling tiles.} \
Space-filling tiles are a small set of 3D shapes whose instances are used to fill a 3D space that is typically planar.
By using a fixed boundary frame, a planar assembly of space-filling tiles can be interlocking, forming a so-called topological interlocking (TI) assembly~\cite{Dyskin-2019-TIMaterials}.
Researchers have shown that a planar TI assembly can be constructed from
all platonic bodies~\cite{Dyskin-2003-TIPlatonicSolids},
convex polyhedra~\cite{Weizmann-2017-BuildingFloors}, and
blocks with curved contact surfaces~\cite{Dyskin-2003-OsteomorphicBlock}.
Recently, researchers have proposed space-filling tiles with more complex shapes, including
Delaunay Lofts tiles~\cite{Subramanian-2019-DelaunayLofts},
generalized Abeille tiles~\cite{Akleman-2020-GeneralizedAbeilleTiles}, and
bi-axial woven tiles~\cite{Krishnamurthy-2020-Bi-AxialWovenTiles, Krishnamurthy-2022-Space-FillingTiling}.
Motivated by the intriguing properties of planar TI assemblies, researchers modeled freeform TI assemblies by lifting planar TI assemblies onto a 3D freeform surface~\cite{Wang-2019-TopoLock, Bejarano-2019-TIDesignFramework, Fallacara-2019-TIForArchitecture}.
However, since the space-filling tiles need to adapt to the curvature of the 3D surface, they do not have identical shapes anymore.
Moreover, the freeform TI assemblies (with a fixed boundary frame) are generally not interlocking~\cite{Wang-2019-TopoLock}.

Compared with the space-filling tiles, our interlocking tiling approach computes a small prototile set whose instances can exactly tile a given 3D finite domain, while ensuring interlocking of the tiling result without relying on a fixed boundary frame.

\begin{figure*}[t]
	\centering
	\includegraphics[width=17.5cm]{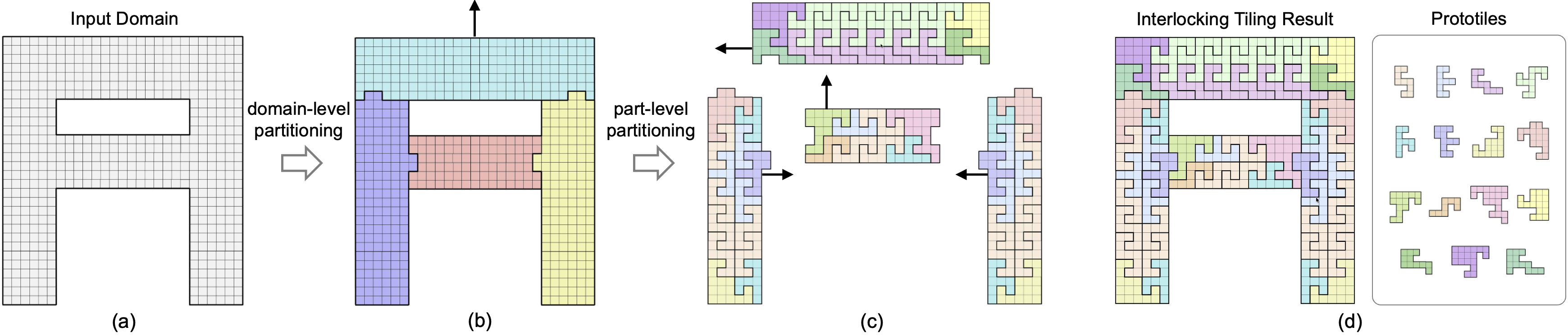}
	\vspace*{-2.0mm}
	\caption{
		Overview of our hierarchical computational approach.
		Given (a) a voxelized input domain, our interlocking tiling approach proceeds in two stages:
		(b) \emph{domain-level partitioning} partitions the domain into interlocking parts such that, once assembled, only one part can be directly taken out;
		(c) \emph{part-level partitioning} further partitions each part into interlocking tiles such that, once assembled, only one tile can be directly taken out.
		(d) The interlocking tiling result  exactly covers the input domain using a small set of prototiles computed by our approach.
	}
	\vspace*{-2.0mm}
	\label{fig:overview}
\end{figure*}

\vspace*{1.0mm}
\noindent
\textbf{Modeling interlocking assemblies.} \
Interlocking is an intriguing property of 3D assemblies where component parts, except the key, are immobilized by their geometric arrangement, making the assemblies structurally stable~\cite{Song-2022-InterlockingSurvey}.
Approaches for modeling interlocking assemblies can be classified as {\em forward} and {\em inverse}, according to whether the shapes of interlocking parts are predefined or computed.

In forward approaches, Zhang and Balkcom~\cite{Zhang-2016-InterlockVoxel} presented a voxel-like interlocking cube set whose instances can be assembled layer-by-layer into various voxelized shapes.
Shih~\cite{Shih-2016-SLBlocks, Shih-2018-SLBlocks, Shih-2020-SLBlockGrammars} presented an SL block whose instances can be assembled following predefined rules to form a special family of large and stable structures.
Sniffen et al.~\cite{Sniffen-2021-FallingIntoPlace} presented a rectilinear-z block set with two blocks whose instances can be used to assemble nearly interlocking planar structures by dropping these instances into place.
Sun et al.~\cite{Sun-2024-StructCurves} demonstrated that various solid-line interlocking structures can be built from instances of two template blocks.

In inverse approaches, when a target shape is given, modeling interlocking assemblies is formulated as a shape decomposition problem, e.g., to design
interlocking puzzles for recreation~\cite{Song-2012-InterCubes, Xin-2011-BurrPuzzles, Tang-2019-3DDissection, Chen-2022-HighLevelPuzzle} or
interlocking assemblies for 3D printing large objects~\cite{Song-2015-Interlock, Yao-2017-InterlockShell}.
When a set of initial parts without joints is given, modeling interlocking assemblies is formulated as a joint planning problem, e.g., for designing interlocking furniture~\cite{Fu-2015-Furniture, Song-2017-ReconfigInterlock}.
In particular, Wang et al.~\cite{Wang-2018-DESIA} developed a unified framework to design interlocking assemblies of different forms by leveraging a graph-based representation.

Forward approaches model interlocking assemblies by using instances of a few predefined interlocking blocks, yet the modeled results can only approximate the target shape.
Inverse approaches model interlocking assemblies that exactly represent the target shape, yet the generated parts typically have distinct shapes.
In this work, we propose a new concept called {\em interlocking tiling}, aiming to combine the complementary strengths of forward and inverse approaches for modeling interlocking assemblies.
Our modeled interlocking assemblies are able to exactly represent the target shape while minimizing the number of unique tile shapes in the results.

\section{Interlocking Tiling Problem}
\label{sec:problem}
To start, we first provide a formal definition of the {\em interlocking tiling} problem by elaborating on its inputs, outputs, and requirements.

\begin{itemize}[leftmargin=*]
	\item
	{\em Inputs.} \
	\begin{enumerate}%
	\vspace*{1.0mm}
	\item
	{\em 3D input domain $D$.} \
	To make the problem tractable, we assume that the input 3D finite domain is represented by a voxelized shape $D$ with $M$ voxels.
	Internal cavities are allowed in the domain.

	\vspace*{1.0mm}
	\item
	{\em Target number of tiles $N_T^{\mathrm{target}}$. } \
	We allow users to set a target number of tiles in the tiling result, denoted as $N_T^{\mathrm{target}}$.
	\end{enumerate}

	\vspace*{1.0mm}
	\item
	{\em Outputs.} \
	\begin{enumerate}%
		\vspace*{1.0mm}
		\item
		{\em Prototile set.} \
		The set of prototiles is denoted as $\{ t_k \}_{k=1}^K$, where $t_k$ is a prototile and $K$ is the total number of prototiles.

		\vspace*{1.0mm}
		\item
		{\em Number of tiles and parts.} \
		The number of tiles is denoted as $N_T$, and the number of parts is denoted as $N_P$.

		\vspace*{1.0mm}
		\item
		{\em Number of instances of each prototile.} \
		The number of instances of each prototile $t_k$ is denoted as $n_k$, which should satisfy $\sum_{k=1}^K n_k = N_T$.

		\vspace*{1.0mm}
		\item
		{\em Transformation of each prototile instance.} \
		The $j$th instance of prototile $t_k$ is denoted as tile $t_{k, j}$, $j \in \{1, 2, \ldots, n_k\}$.
		The placement location and orientation of each tile $t_{k, j}$ in domain $D$ are represented by a transformation denoted as $T_{k, j}$.

		\vspace*{1.0mm}
		\item
		{\em Assembly plan.} \
		Our approach also computes an assembly plan for assembling the computed tiles into the interlocking tiling result.
	\end{enumerate}

	\vspace*{1.0mm}
	\item
	{\em Requirements.} \
	\begin{enumerate}%
		\vspace*{1.0mm}
		\item
		{\em Exact domain coverage.} \
		The computed tiles $\{ t_{k, j} \}$ should exactly cover domain $D$ without gaps or overlaps.

		\vspace*{1.0mm}
		\item
		{\em Domain-level interlocking.} \
		At the domain level, the interlocking tiling result can be viewed as an interlocking assembly of parts, where only one part, called the key part, can be directly taken out once assembled.

		\vspace*{1.0mm}
		\item
		{\em Part-level interlocking.} \
		At the part level, each part is an interlocking subassembly of tiles, where only one tile, called the key tile, can be directly taken out once assembled.

		\vspace*{1.0mm}
		\item
		{\em Minimizing the number of prototiles.} \
		The number of prototiles, $K$, for tiling domain $D$ should be minimized.

		\vspace*{1.0mm}
		\item
		{\em Matching the target number of tiles.} \
		The number of tiles, $N_T$, should be as close as possible to the target $N_T^{\mathrm{target}}$.

		\vspace*{1.0mm}
		\item
		{\em Disassemblability.} \
		The interlocking tiling result can be completely disassembled into individual tiles.
	\end{enumerate}
\end{itemize}

\section{Hierarchical Computational Approach}
\label{sec:approach}
In this paper, we propose a hierarchical approach to solve the interlocking tiling problem.
Our hierarchical approach has two stages:
(i) {\em domain-level partitioning}, which partitions the input domain into parts such that these parts can be assembled into an interlocking assembly (Figure~\ref{fig:overview}(a\&b)); and
(ii) {\em part-level partitioning}, which further partitions each part into tiles such that the tiles within each part form an interlocking subassembly (Figure~\ref{fig:overview}(b\&c)) and the entire set of tiles has a small number of distinct shapes (Figure~\ref{fig:overview}(d)).

\begin{figure}[t]
	\centering
	\includegraphics[width=8.35cm]{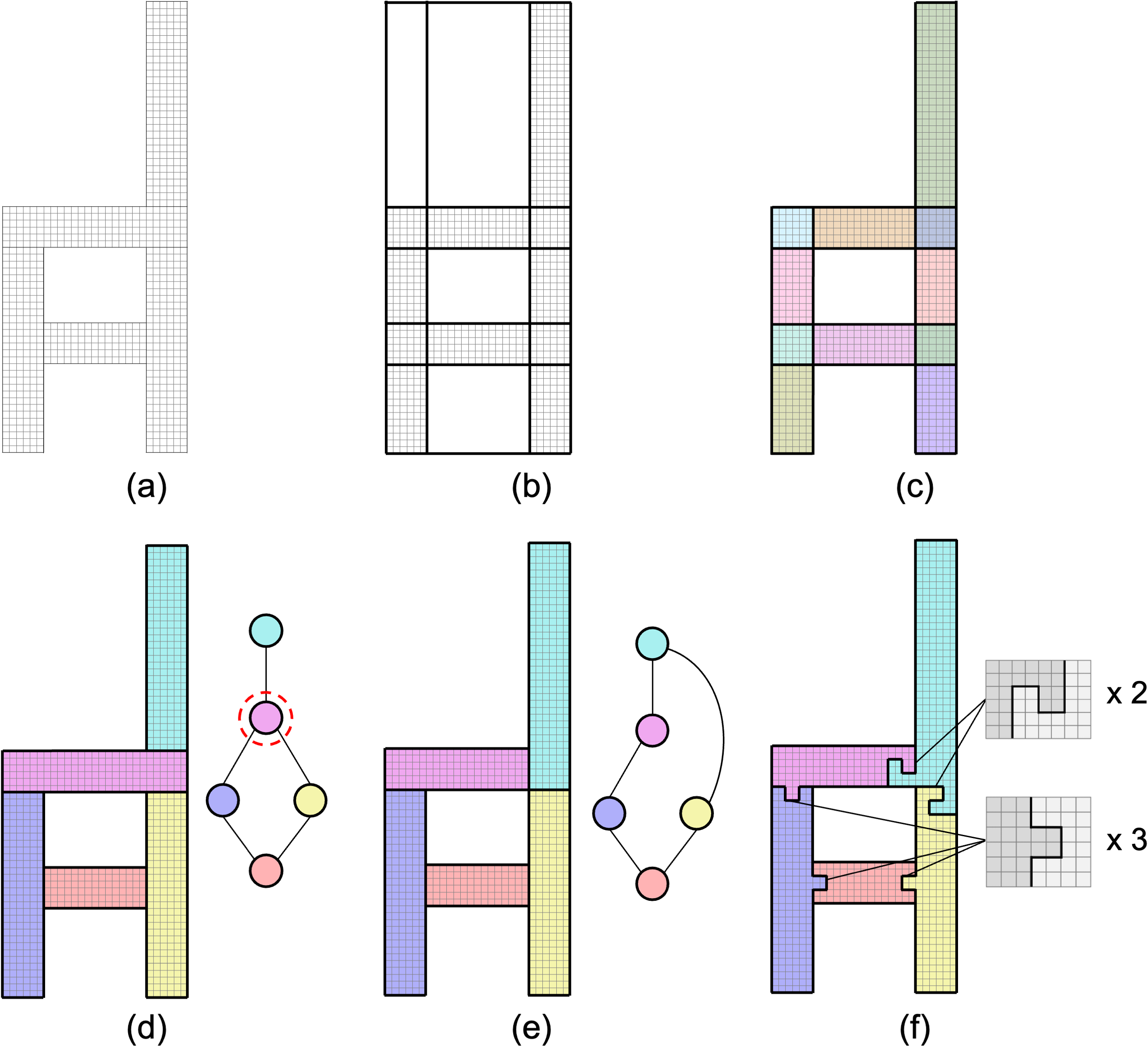}
	\caption{
		At the domain-level partitioning stage, we partition the input domain into parts, such that these parts can be assembled into an interlocking assembly.
		Given (a) an input domain, we first detect all (b) valid axis-aligned cutting planes, and apply these cutting planes to induce (c) a cell decomposition.
		Next, we merge the cells into parts to form (d\&e) candidate partitions of the input domain, where the partition in (d) is not valid since its part-graph has a cut point.
		Finally, we plan (f) inter-part joints to achieve domain-level interlocking while encouraging joint reuse.
	}
	\vspace*{-2.0mm}
	\label{fig:approach_domain}
\end{figure}

\subsection{Domain-level Partitioning}
Given a voxelized 3D finite domain, the domain-level partitioning stage aims to partition the input domain into parts $\{ P_i \}$.
This partitioning serves two purposes:
(1) it ensures the domain-level interlocking property; and
(2) it decomposes a complex domain into parts with simple shapes, which facilitates the subsequent part-level partitioning for generating tiles that have a small number of distinct shapes.
This stage consists of four steps:

\begin{enumerate}[leftmargin=*]
	\item
	\emph{Splitting the domain into cells.}
	We first decompose the input domain into cells using axis-aligned cutting planes.
	These planes are placed along the boundaries of the domain and extended across the entire domain, thereby partitioning it into box-shaped cells.
	Collectively, the cutting planes induce a cell decomposition of the input domain; see Figure~\ref{fig:approach_domain}(a-c) for an example.

	\vspace*{1mm}
	\item
	\emph{Merging cells into parts.}
	We then merge the cells into parts to generate candidate partitions of the input domain.
	To construct each part, we first randomly select a cell as its initial cell and then iteratively merge it with a neighboring cell, provided that the merged result remains box-shaped.
	The merging process terminates when the maximum number of cells (denoted by $m_\text{cell}$) allowed for a part is reached or when no valid neighboring cell can be merged.
	Once a part is constructed, its cells are removed from further consideration, and the same procedure is repeated to construct the next part until all cells in the input domain are assigned to parts.
	We repeat this randomized algorithm to generate multiple candidate partitions; see Figure~\ref{fig:approach_domain}(d\&e) for two examples, where $m_\text{cell} = 3$.
	By default, each part is box-shaped.
	However, users  are allowed to manually merge multiple parts to create more complex geometries, such as a plate with a hole or a cavity, as shown in Figure~\ref{fig:teaser}.

	\vspace*{1mm}
	\item
	\emph{Selecting a partition.}
	For each candidate partition, we model its part connections with a part graph, in which nodes represent parts and edges represent contacts between parts.
	A candidate partition is valid only if its part graph has no cut point, since a cut point prevents an interlocking assembly regardless of the joint design~\cite{Wang-2018-DESIA}.
	Figure~\ref{fig:approach_domain}(d\&e) shows an invalid and a valid partition, respectively.
	Among the valid candidate partitions, we prioritize those with fewer parts, as they are generally easier to assemble, and present them to the user for selection.

	\vspace*{1mm}
	\item
	{\em Planning joints.}\
	We then plan inter-part joints to achieve domain-level interlocking.
	Specifically, joints are introduced by reassigning voxels between adjacent parts.
	We extend the algorithm in~\cite{Fu-2015-Furniture} to plan joints by prioritizing generating as many identical joint geometries as possible.
	Figure~\ref{fig:approach_domain}(f) shows an example in which five joints fall into two classes of geometries.
	This modification facilitates the subsequent part-level partitioning by leaving more coherent space within each part.
\end{enumerate}

\subsection{Part-level Partitioning}
After domain-level partitioning, the input domain has been partitioned into a set of parts $\{ P_i \}$ that interlock with one another. In the part-level partitioning stage, we further partition each part into tiles, with two goals: (1) ensuring that the computed tiles within each part form an interlocking subassembly; and (2) minimizing the number of prototiles globally over the entire domain.
Part-level partitioning is the key step for achieving high prototile reuse.

Compared with the domain-level partitioning, part-level partitioning is significantly more challenging due to its two inherently conflicting goals.
On the one hand, achieving interlocking requires tiles with complex and distinct shapes so that they can effectively block one another.
On the other hand, achieving high prototile reuse favors tiles with simple and identical shapes.
These two objectives are fundamentally at odds, making it difficult to simultaneously achieve interlocking and high prototile reuse.

To achieve the first goal, Wang et al.~\cite{Wang-2018-DESIA} proposed a Directional Blocking Graphs (DBGs)-based approach for designing interlocking assemblies, where blocking relations required within interlocking assemblies are encoded in the computed DBGs and subsequently realized in geometry space.
However, since the DBGs in their approach are computed via random search, they often exhibit irregular structures.
Such irregularity in graph space tends to translate into irregular geometric realizations, resulting in interlocking tiles that are almost all distinct in shape (i.e., failing to achieve the second goal).
Based on this observation, our key idea is to develop a class of special DBGs with regular structures to encode necessary blocking relations within interlocking assemblies in graph space, and then use it to guide the generation of interlocking tiles that have a small number of distinct shapes.

In the following, Section~\ref{sec:graph} introduces Interlocking Tiling Directional Blocking Graphs (IT-DBGs), a family of graphs with regular structures, to represent blocking relations in interlocking tilings.
Building upon this representation, Section~\ref{sec:partition} presents our part-level partitioning approach that leverages IT-DBGs to guide the partitioning of each part into interlocking tiles while minimizing the number of prototiles.

\section{Interlocking Tiling DBGs}
\label{sec:graph}
In this section, we first review base DBGs~\cite{Wilson-1992-GeometricAssemblyPlanning} and how they can be used to efficiently test interlocking~\cite{Wang-2018-DESIA}.
We then present \emph{Interlocking Tiling DBGs}, a special family of base DBGs, to represent single-layer interlocking tilings with a regular $n \times m$ grid structure.
Such a graph representation with regular structures encourages the generation of repetitive tile shapes, thereby promoting prototile reuse during part-level partitioning.
Finally, we introduce a graph concatenation scheme that combines multiple sets of IT-DBGs, whose geometric realizations can represent complex 3D shapes.

\emph{Base DBGs.}
Given a 3D assembly consisting of $n$ pieces $\{p_1, p_2, \ldots, p_n\}$, blocking relations can be encoded using DBGs.
In a DBG denoted by $G(d)$ where $d \in \{\pm x,\pm y,\pm z\}$, each node represents a piece and each directed edge from $p_i$ to $p_j$ indicates that $p_i$ blocks $p_j$ from moving along direction $d$.
Since blocking reverses when the moving direction is negated, $G(-d)$ is simply the edge-reversed graph of $G(d)$.
Therefore, it suffices to analyze the base set $\{G(+x),G(+y),G(+z)\}$ (i.e., the base DBGs) to cover all six directions in a 3D domain.

\emph{Interlocking Testing using DBGs.}
According to~\cite{Wang-2018-DESIA}, testing whether a 3D assembly is interlocking is equivalent to checking a graph condition on the assembly's base DBGs.
Specifically, an assembly with at least three pieces is interlocking if,
for each base direction $d \in \{+x,+y,+z\}$, the corresponding $G(d)$ satisfies one of the following conditions:
\begin{enumerate}[label=(\arabic*)]
	\item $G(d)$ is strongly connected; or
	\item $G(d)$ has exactly two strongly connected components (SCCs), one of which has a single piece that is identical across all DBGs.
\end{enumerate}

\begin{wrapfigure}{r}{0.45\columnwidth}
	\centering
	\vspace{-10pt} %
  	\hspace{-15pt}
	\includegraphics[width=0.43\columnwidth]{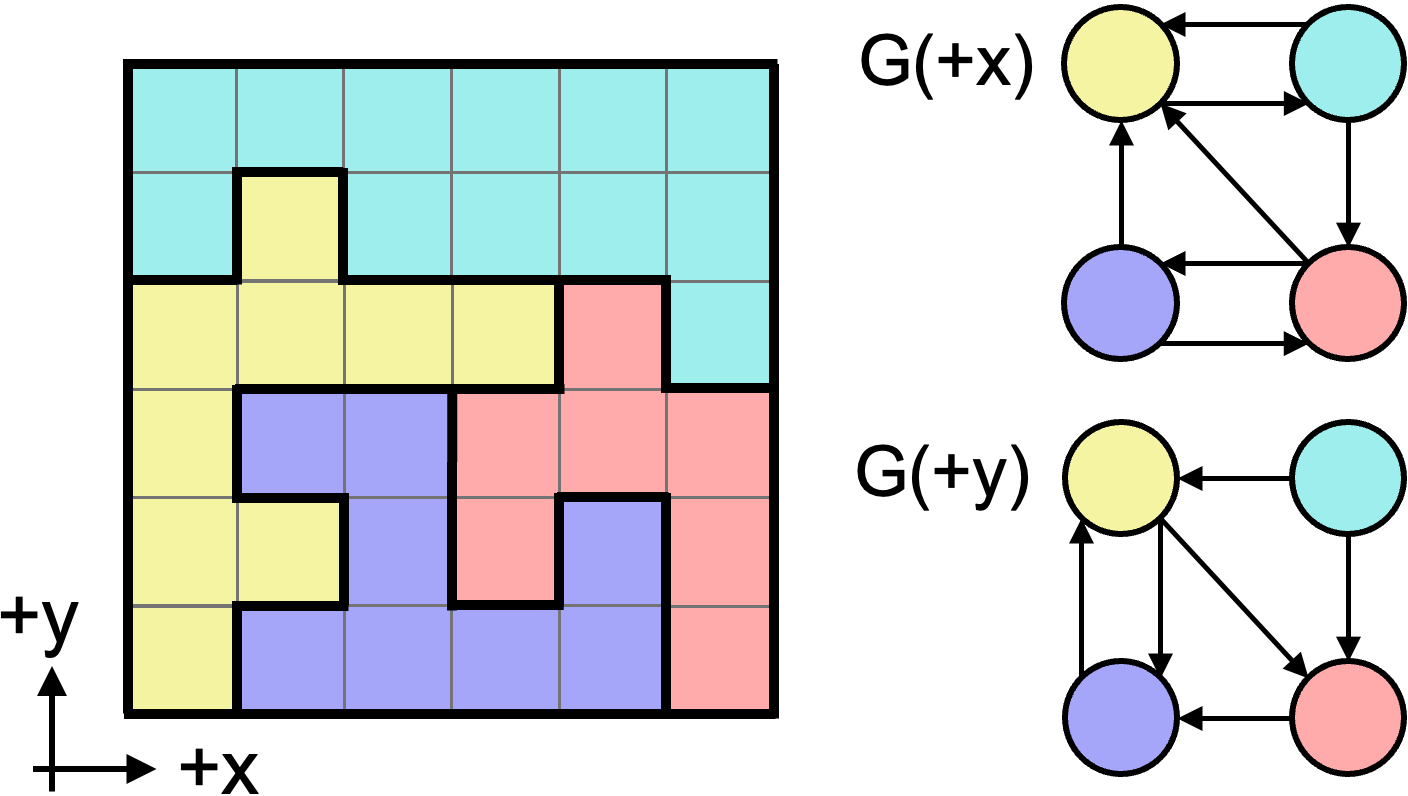}
	\vspace{-10pt}
\end{wrapfigure}
The inset figure shows a 2D interlocking assembly with 4 pieces, where the key piece is in cyan.
For its two base DBGs, $G(+x)$ is strongly connected, and $G(+y)$ has exactly two SCCs, where one SCC contains the key piece and the other SCC contains the remaining three pieces.

\subsection{Definition and Properties}

We define IT-DBGs, denoted by $\mathcal{G}$, as a special family of base DBGs along the three positive axes:
\[
\mathcal{G} \;=\; \{G^\mathrm{IT}(+x),\,G^\mathrm{IT}(+y),\,G^\mathrm{IT}(+z)\}.
\]
Each base DBG is a directed graph $G^\mathrm{IT}(d)=(V, E(d))$, where $V$ is the node set shared by the three base DBGs and $E(d)$ is the directed edge set encoding blocking relations between tiles along direction $d$, where  $d \in \{+x, +y, +z\}$.

For all base DBGs, we arrange the node set $V$ on a regular $n\times m$ grid with $n\ge 2$ rows and $m\ge 3$ columns:
\[
V=\{V_{i,j}\mid i \in [1, n], j \in [1, m]\}.
\]
Each node $V_{i,j}$ corresponds to the tile located at row $i$ and column $j$.
Let $V^{\mathrm{key}}$ denote the key node (i.e., the node corresponding to the key tile).
The row--column coordinate system of the IT-DBGs is defined with respect to the key, meaning that the key always lies in the first row ($i=1$).

\begin{figure}[!t]
	\centering
	\includegraphics[width=8.30cm]{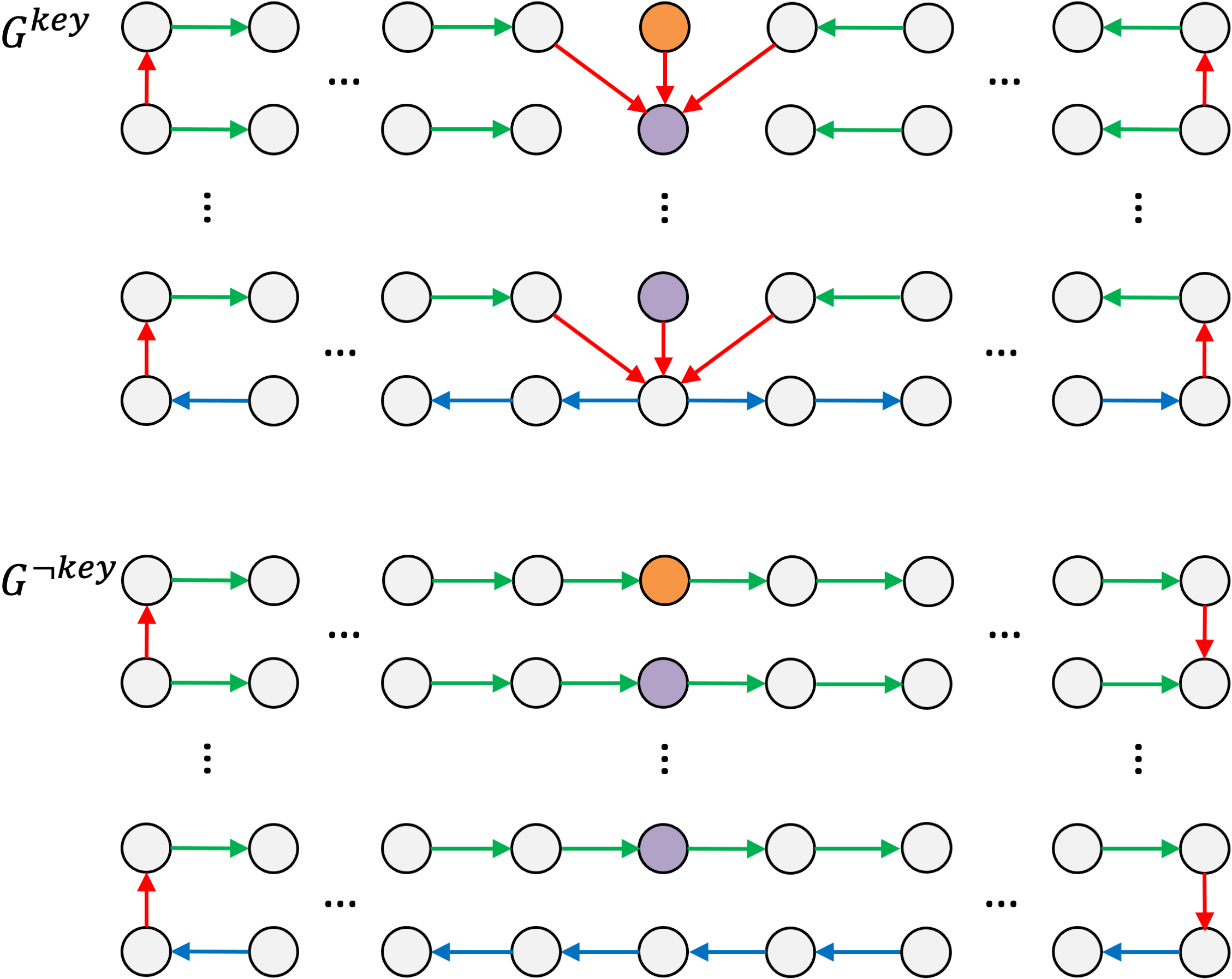}
	\vspace*{-1.0mm}
	\caption{
		We propose interlocking tiling DBGs to guide part-level partitioning.
		In 3D space, each set of IT-DBGs consists of one key-bearing base graph $G^{\text{key}}$ and two non-key base graphs that share an identical structure.
		Edge colors indicate the three edge groups:
		edges within the rows above the last one (green),
		edges within the last row (blue), and
		edges between adjacent rows (red).
		The orange node is the key node, while each purple node is one that takes over as the key once the rows above it have been removed.
	}
	\vspace*{-2.0mm}
	\label{fig:approach_graph}
\end{figure}

Among the three base DBGs, we call the one defined along the key-removal direction the \emph{key-bearing} base DBG and denote it by $G^{\mathrm{key}}$; the remaining two are designed to be identical, referred to as the \emph{non-key} base DBGs and are denoted by $G^{\neg\mathrm{key}}$.
We next specify the edge sets $E^{\mathrm{key}} = E^{\mathrm{key}}_{1}\ \cup E^{\mathrm{key}}_{2}\ \cup E^{\mathrm{key}}_{3}$ and $E^{\neg\mathrm{key}} = E^{\neg\mathrm{key}}_{1}\ \cup\ E^{\neg\mathrm{key}}_{2}\ \cup\ E^{\neg\mathrm{key}}_{3}$ for the key-bearing and non-key base DBGs, respectively, which are defined as follows:
\begin{enumerate}[label=(\arabic*), leftmargin=*]
	\item
	\emph{Edges in non-last rows:}
	$E^{\mathrm{key}}_{1}$ ($E^{\neg\mathrm{key}}_{1}$) contains all edges within rows $1 \le i < n$ of $G^{\mathrm{key}}$ ($G^{\neg\mathrm{key}}$); see the green edges in Figure~\ref{fig:approach_graph}.

	\item
	\emph{Edges in the last row:}
	$E^{\mathrm{key}}_{2}$ ($E^{\neg\mathrm{key}}_{2}$) contains all edges within row $i=n$ of $G^{\mathrm{key}}$ ($G^{\neg\mathrm{key}}$); see the blue edges in Figure~\ref{fig:approach_graph}.

	\item
	\emph{Inter-row edges:}
	$E^{\mathrm{key}}_{3}$ ($E^{\neg\mathrm{key}}_{3}$) contains all edges between adjacent rows of $G^{\mathrm{key}}$ ($G^{\neg\mathrm{key}}$); see the red edges in Figure~\ref{fig:approach_graph}.
\end{enumerate}
These edge sets are defined to support the subsequent part-level partitioning stage, where the blocking relations encoded within each edge set are geometrically realized sequentially.

\emph{Graph properties.}
Our proposed IT-DBGs are designed to be \emph{interlocking}, \emph{recursive}, and \emph{non-redundant}:
\begin{enumerate}[label=(\arabic*), leftmargin=*]
	\item
	\emph{Globally interlocking.} \
	IT-DBGs satisfy the interlocking condition in graph space since $G^{\neg\mathrm{key}}$ is strongly connected and $G^{\mathrm{key}}$ has exactly two strongly connected components; see again Figure~\ref{fig:approach_graph}.
	In $G^{\neg\mathrm{key}}$, each non-last row forms a cycle with the last row through columns $1$ and $m$, so every node is mutually reachable.
	In $G^{\mathrm{key}}$, the key column splits the grid into a left and a right side: the left side repeats the cycle structure of $G^{\neg\mathrm{key}}$ with the key node excluded, and the right side is its mirror image.
	The key node is therefore the only node lying on no cycle, and forms an SCC by itself while all remaining nodes form the other.

	\vspace*{1.0mm}
	\item
	\emph{Row-wise recursive interlocking.} \
	If IT-DBGs have more than three rows, removing the first-row nodes, i.e., $\{V_{1,j}\}$, from both $G^{\mathrm{key}}$ and $G^{\neg\mathrm{key}}$ still preserves the interlocking condition.
	This property enhances structural stability during the progressive (dis)assembly process.
	Figure~\ref{fig:approach_graph} highlights two nodes in purple that can serve as the key after removing the top rows.

	\vspace*{1.0mm}

	\item
	\emph{No redundant blocking relation.} \
	Removing any edge from an IT-DBG causes a violation of either Property~1 or Property~2.
	Thus, every encoded blocking relation is necessary in our setting,
	which encourages simple geometric realizations and, in turn, promotes prototile reuse.
\end{enumerate}

\subsection{Concatenating Multiple Sets of IT-DBGs}
\label{subsec:concatenate_graph}
\begin{figure*}[!t]
	\centering
	\includegraphics[width=16.0cm]{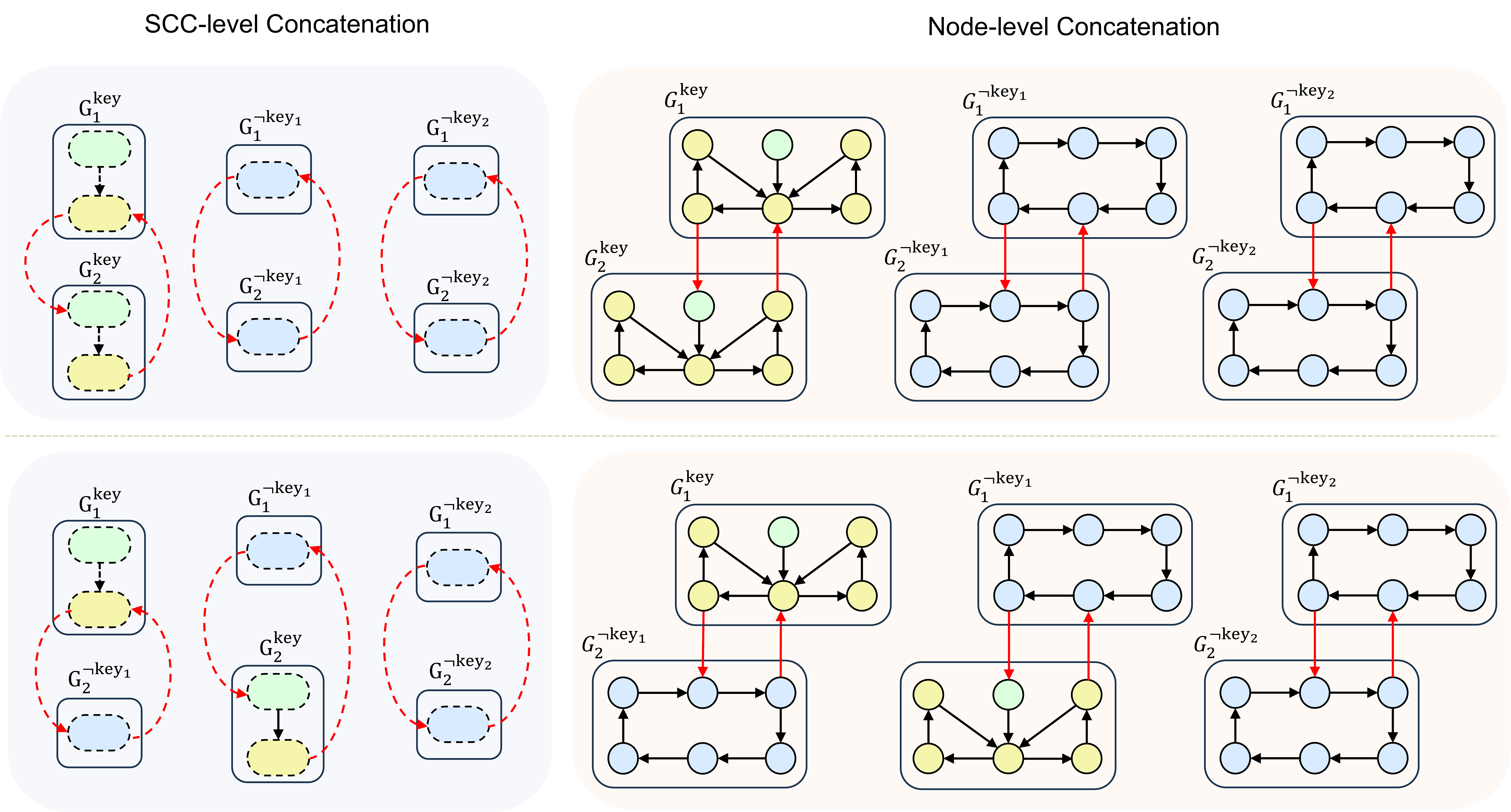}
	\caption{
		Concatenating two sets of IT-DBGs $\mathcal{G}_1$ and $\mathcal{G}_2$, whose key-removal directions are (top) identical or (bottom) different.
		In both cases, (left) we concatenate each pair of corresponding base graphs by introducing SCC-level edges (red dashed edges) between strongly connected components (dashed ellipses),
		and then (right) implement these edges via node-level connections (red solid edges).
	}
	\label{fig:approach_concatenation}
\end{figure*}

A single set of IT-DBGs yields a single-layer geometric realization with a regular $n \times m$ grid structure.
To represent complex 3D shapes, we need to concatenate multiple sets of IT-DBGs while ensuring that the resulting graphs still satisfy the interlocking condition.
Recall that each set of IT-DBGs consists of one key-bearing base graph $G^{\text{key}}$ and two non-key base graphs that share an identical structure, denoted by $G^{\neg \text{key}}$.
In the following, we use $G^{\neg \text{key}_1}$ and $G^{\neg \text{key}_2}$ to distinguish the two non-key base graphs when necessary.
$G^{\text{key}}$ contains two strongly connected components: the key component $C^{\text{key}}$ (see the green dashed ellipses in Figure~\ref{fig:approach_concatenation} (left)) and the remaining component $C^{\text{rest}}$ (see the yellow dashed ellipses in Figure~\ref{fig:approach_concatenation} (left)),
whereas each $G^{\neg \text{key}_j}$, $j = 1, 2$, forms a single SCC (see the blue dashed ellipses in Figure~\ref{fig:approach_concatenation} (left)).

When concatenating two sets of IT-DBGs $\mathcal{G}_1$ and $\mathcal{G}_2$, we retain the key of $\mathcal{G}_1$ without loss of generality.
To preserve the interlocking condition, we connect their corresponding base graphs so that all nodes except this key form a single SCC along
its key-removal direction, while all nodes form a single SCC along each of the other two directions.
We distinguish two cases according to whether the key-removal directions of $\mathcal{G}_1$ and $\mathcal{G}_2$ are identical:
\begin{enumerate}[label=(\arabic*), leftmargin=*]
	\item
	\emph{Identical key-removal directions.} \
	We add two edges,
	$C^{\text{rest}}_1 \rightarrow C^{\text{key}}_2$ and
	$C^{\text{rest}}_2 \rightarrow C^{\text{rest}}_1$,
	which together with the existing edge
	$C^{\text{key}}_2 \rightarrow C^{\text{rest}}_2$,
	form a directed cycle
	$C^{\text{rest}}_1 \rightarrow C^{\text{key}}_2
	\rightarrow C^{\text{rest}}_2 \rightarrow C^{\text{rest}}_1$.
	This merges the three components into one SCC while leaving
	$C^{\text{key}}_1$ as a single SCC.
	Along each of the other two directions, we add oppositely directed edges between
	$G_1^{\neg \text{key}_j}$ and $G_2^{\neg \text{key}_j}$,  $j=1,2$,
	making the concatenated base graph strongly connected.
	See Figure~\ref{fig:approach_concatenation} (top) for an illustration.

	\vspace*{1mm}
	\item
	\emph{Different key-removal directions.} \
	Along the key-removal direction of $\mathcal{G}_1$, we add a pair
	of oppositely directed edges between $C^{\text{rest}}_1$ and
	$G_2^{\neg \text{key}_1}$, merging all nodes except the retained key
	into one SCC.
	Along the key-removal direction of $\mathcal{G}_2$, we add two edges,
	$G_1^{\neg \text{key}_1} \rightarrow C^{\text{key}}_2$ and
	$C^{\text{rest}}_2 \rightarrow G_1^{\neg \text{key}_1} $,
	which together with the existing edge
	$C^{\text{key}}_2 \rightarrow C^{\text{rest}}_2$,
	form a directed cycle
	$G_1^{\neg \text{key}_1}  \rightarrow C^{\text{key}}_2
	\rightarrow C^{\text{rest}}_2 \rightarrow G_1^{\neg \text{key}_1} $,
	making the concatenated base graph strongly connected.
	Along the remaining direction, we add a pair of oppositely
	directed edges between $G_1^{\neg \text{key}_2} $ and
	$G_2^{\neg \text{key}_2} $, making the concatenated base graph
	strongly connected.
	See Figure~\ref{fig:approach_concatenation} (bottom) for an illustration.
\end{enumerate}

In practice, each SCC-level connection (red dashed edges in Figure~\ref{fig:approach_concatenation} (left)) is realized by introducing a minimal number of node-level edges (red solid edges in Figure~\ref{fig:approach_concatenation} (right)),
typically between the spatially closest pairs of nodes, to facilitate geometric realization.
Concatenating multiple sets of IT-DBGs can be done by iteratively applying the above graph concatenation process.

\section{Part-level Partitioning Guided by IT-DBGs}
\label{sec:partition}
In this section, we introduce our part-level partitioning approach, which generates interlocking tiles on a part domain by geometrically realizing blocking relations encoded in IT-DBGs.
We first present an approach that assumes the input part is box-shaped (with one or more small joints), and then extend the approach to handle non-box-shaped parts.

Our part-level partitioning approach consists of four steps:

\begin{figure}[!t]
	\centering
	\includegraphics[width=8.50cm]{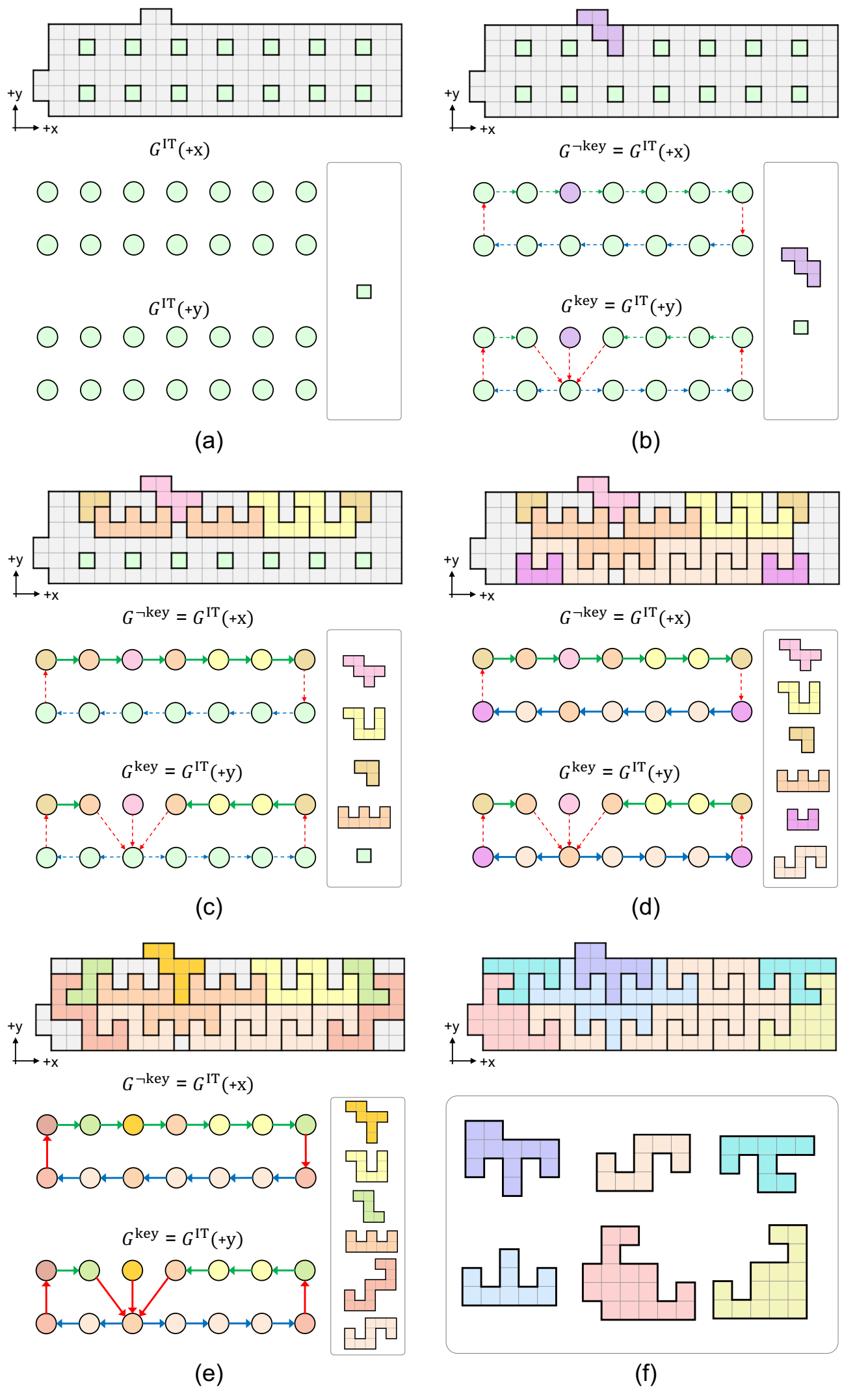}
	\vspace*{-1.5mm}
	\caption{Part-level partitioning guided by IT-DBGs.
		Given a part domain with an $n \times m$ layout, (a) we first initialize tiles together with the node set of the target IT-DBGs.
		(b) Next, we identify the key-removal direction and construct the key tile,
		which determines $G^{\mathrm{key}}$($G^{\neg\mathrm{key}}$) in the target IT-DBGs and the necessary blocking relations for being interlocking (see the dashed edges).
		(c--e) Next we geometrically realize the blocking relations (see the solid edges) in the target IT-DBGs while encouraging prototile reuse.
		(f) Finally, we iteratively assign unoccupied voxels to adjacent tiles.
	}
	\vspace*{-4.0mm}
	\label{fig:approach_part}
\end{figure}

\begin{figure*}[!t]
	\centering
	\includegraphics[width=18.0cm]{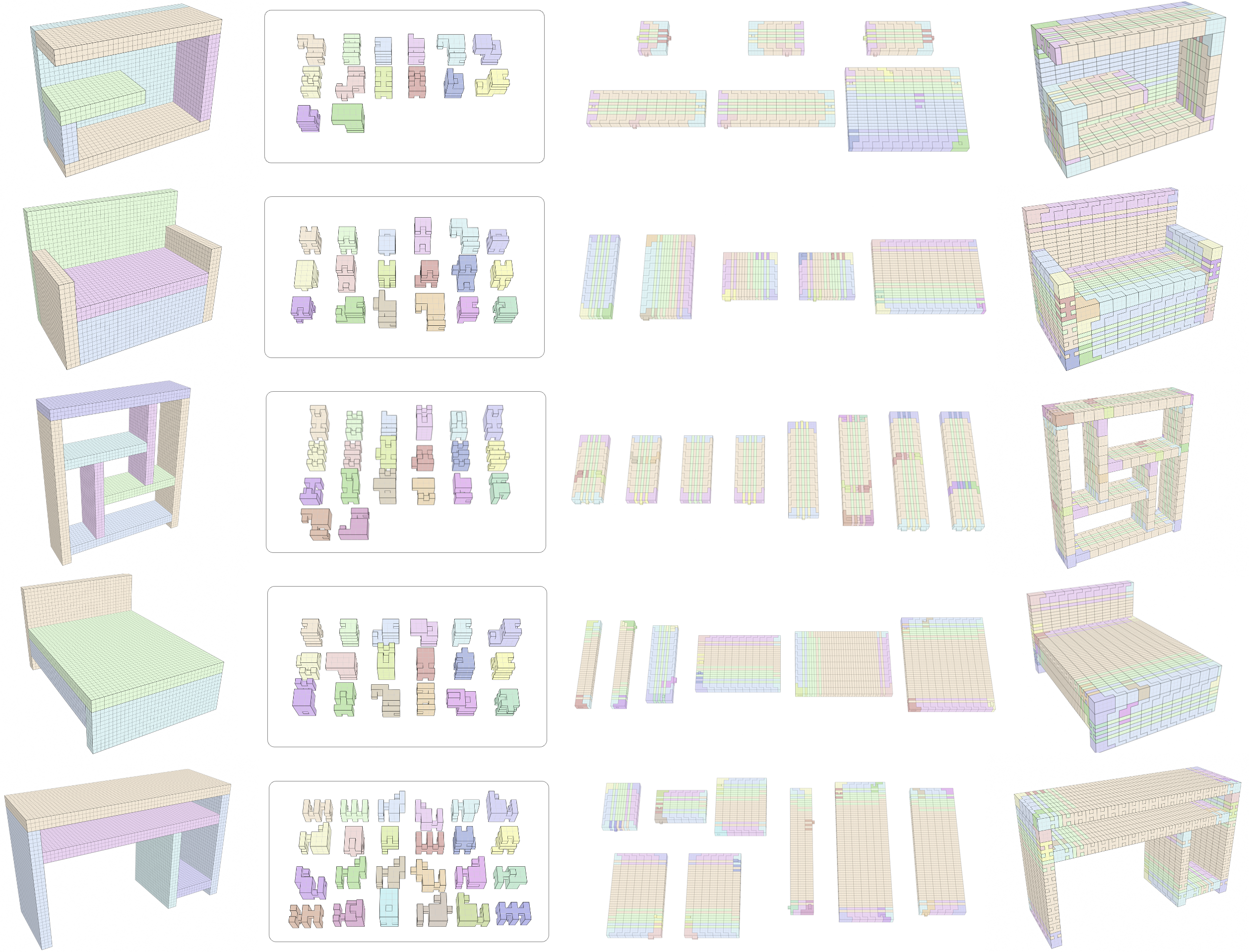}
	\caption{Our two-level partitioning approach is able to generate interlocking tiling results on various spatial domains while achieving high prototile reuse.
	}
	\label{fig:result_shape}
\end{figure*}

\begin{enumerate}[label=(\arabic*), leftmargin=*]
	\vspace*{1.0mm}
	\item
	\emph{Initializing tiles.}
	To facilitate the generation of repetitive tile shapes,
	we initialize tiles in a regular pattern by uniformly sampling seed voxels on a regular integer lattice with a stride vector $\mathbf{s} = (s_x, s_y, s_z)$, where $s_x, s_y, s_z\ge 3$ to ensure sufficient space among initial tiles for realizing blocking relations.
	A voxel at grid index $(i, j, k)$ is selected as a seed if it lies inside the input domain and satisfies
	$i \equiv 0 \pmod{s_x}$,
	$j \equiv 0 \pmod{s_y}$, and
	$k \equiv 0 \pmod{s_z}$.
	Each selected seed voxel initializes a tile as well as the corresponding node in the target IT-DBGs; see Figure~\ref{fig:approach_part}(a).
	The stride vector $\mathbf{s}$ control the number of tiles, and is adjusted according to the user-specified target number of tiles $N_T^{\mathrm{target}}$:
	their values are chosen such that the number of seeds placed in each part $P_i$ is as close as possible to its target tile count $\frac{M_{P_i}}{M} \times N_T^{\mathrm{target}}$,
	where $M_{P_i}$ denotes the number of voxels in $P_i$ and $M$ is the total number of voxels in the domain.

	\vspace*{1.0mm}
	\item
	\emph{Constructing the key tile.}\
	To ensure the two-level interlocking property, the key tile within each part must remain non-removable until the domain-level key part has been removed.
	We therefore construct the key tile so that it can only be \emph{externally blocked} by neighboring parts.
	Once the key-removal direction is chosen, we can identify the key-bearing base graph $G^{\mathrm{key}}$ and the non-key base graph $G^{\neg\mathrm{key}}$ in the target IT-DBGs.
	In Figure~\ref{fig:approach_part}(b), we construct the key tile (see the purple tile) with the key-removal direction $+y$, which determines $G^{\mathrm{key}}=G^\mathrm{IT}(+y)$ and $G^{\neg\mathrm{key}}=G^\mathrm{IT}(+x)$, along with the necessary blocking relations for being interlocking (see the dashed edges) in the target IT-DBG.

	\vspace*{1.0mm}
	\item
	\emph{Geometrically realizing blocking relations.}
	We geometrically realize the blocking relations in the target IT-DBGs in three steps:
	(i) realizing edge sets $E_1^{\mathrm{key}}$ and $E_1^{\neg\mathrm{key}}$ within non-last rows; see green solid edges in Figure~\ref{fig:approach_part}(c);
	(ii) realizing edge sets $E_2^{\mathrm{key}}$ and $E_2^{\neg\mathrm{key}}$ within the last row; see blue solid edges in Figure~\ref{fig:approach_part}(d); and
	(iii) realizing edge sets $E_3^{\mathrm{key}}$ and $E_3^{\neg\mathrm{key}}$ between rows; see red solid edges in Figure~\ref{fig:approach_part}(e).
	We realize each edge following the approach in~\cite{Wang-2018-DESIA}.
	In detail, to realize an edge from tile $p_i$ to tile $p_j$ in $G(d)$, we first select two anchor voxels $v_i$ and $v_j$,
    where $v_i$ is adjacent to $v_j$ along direction $d$, and then assign $v_i$ to $p_i$ and $v_j$ to $p_j$, respectively.
	Next, we find a shortest path from $v_i$ to $p_i$ (from $v_j$ to $p_j$) and assign all the voxels on the path to tile $p_i$ (tile $p_j$).
	To encourage prototile reuse, we prioritize realizations that lead to tiles whose shapes match those of existing tiles.
	After realizing each edge, we check whether the resulting assembly is deadlocking.
	If deadlocking, we backtrack and try alternative voxel assignments.

	\vspace*{1.0mm}
	\item
	\emph{Assigning unoccupied voxels.}
	After geometrically realizing all the edges in the target IT-DBGs, some voxels in the part domain may remain unoccupied (Figure~\ref{fig:approach_part}(e)).
	We iteratively assign these voxels to their neighboring tiles while encouraging prototile reuse using an approach similar to~\cite{Chen-2025-InverseTiling}.
	In detail, at each iteration, we select a batch of candidate tiles of identical shape and grow each of them by one neighboring unoccupied voxel, keeping their shapes identical after growth.
	If not all selected tiles can grow together while remaining congruent, we grow the largest possible subset identically and leave the remaining tiles unchanged, which increases the number of prototiles $K$ by one.
	Each assignment is accepted only if the tiling result remains interlocking.
	After all unoccupied voxels are assigned, we compute a feasible plan that disassembles the interlocking tiling result into individual tiles.

\end{enumerate}

\begin{figure}[!t]
	\centering
	\includegraphics[width=8.50cm]{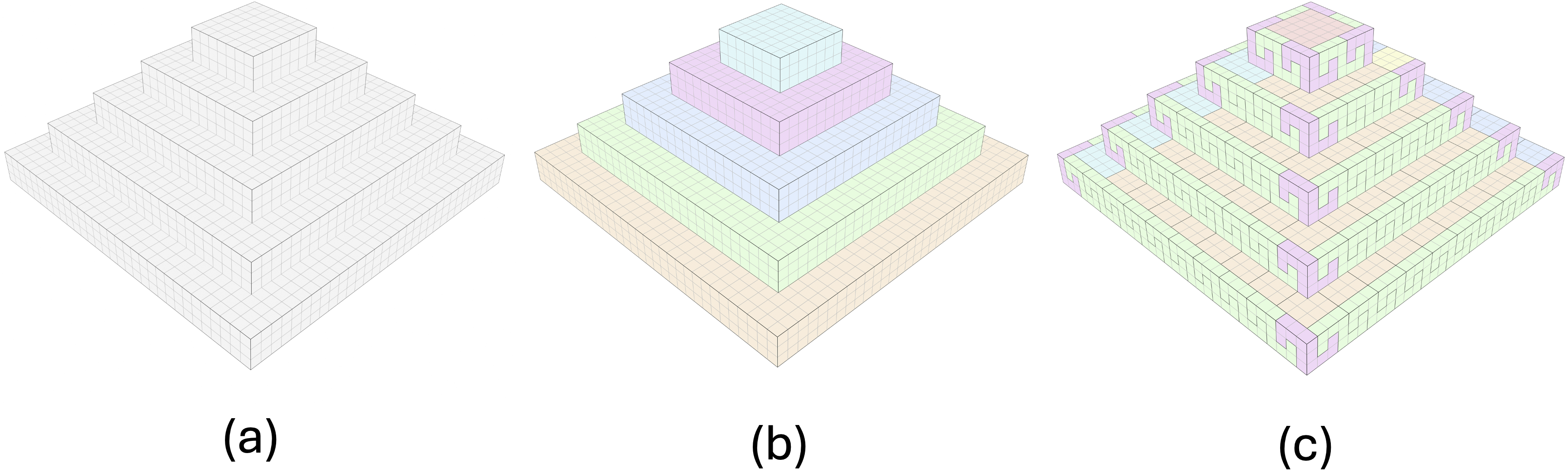}
	\vspace*{-1.5mm}
	\caption{
	Given (a) a non-box-shaped part, we (b) slice it and treat each slice as a subpart.
	We initialize a regular seed layout within the maximum inscribed cuboid of each subpart,
	then geometrically realize the blocking relations within each set of IT-DBGs as well as those between different sets of IT-DBGs,
	and assign the remaining unoccupied voxels to (c) obtain an interlocking tiling result.
	}
	\vspace*{-4.0mm}
	\label{fig:approach_pyramid}
\end{figure}

{\em Extension to non-box-shaped parts.} \
The above approach assumes that the input part is box-shaped and thus block relations within the part can be represented by one set of IT-DBGs.
To extend the approach to non-box-shaped parts, we approximate the input part with a few box-shaped subparts, represented the block relations within each subpart using one set of IT-DBGs, and concatenate these sets of IT-DBGs using the approach in Section~\ref{subsec:concatenate_graph} to ensure the resulting IT-DBGs still satisfy the interlocking condition.

In detail, we make the following changes to the above approach.
First, in Step~1 (i.e., initializing tiles), we slice the input part and treat each slice with thickness $h$ as a subpart.
Within each subpart, we place an $n \times m$ layout of seeds in its maximum inscribed cuboid using strides $s_x$ and $s_y$, so that each subpart corresponds to one set of IT-DBGs.
Together, $s_x, s_y$ and thickness $h$ are adjusted according to the user-specified target number of tiles $N_T^{\mathrm{target}}$.
In Step~3 (i.e., geometrically realizing blocking relations), we first geometrically realize the blocking relations within each subpart represented by each set of IT-DBGs, and then realize the blocking relations among subparts represented by the new edges introduced when concatenating multiple sets of IT-DBGs (i.e., red solid edges in Figure~\ref{fig:approach_concatenation}).
Figure~\ref{fig:approach_pyramid} shows an example with five subparts and eight prototiles.

\section{Results}
\label{sec:result}

\begin{figure*}[!t]
	\centering
	\includegraphics[width=18.0cm]{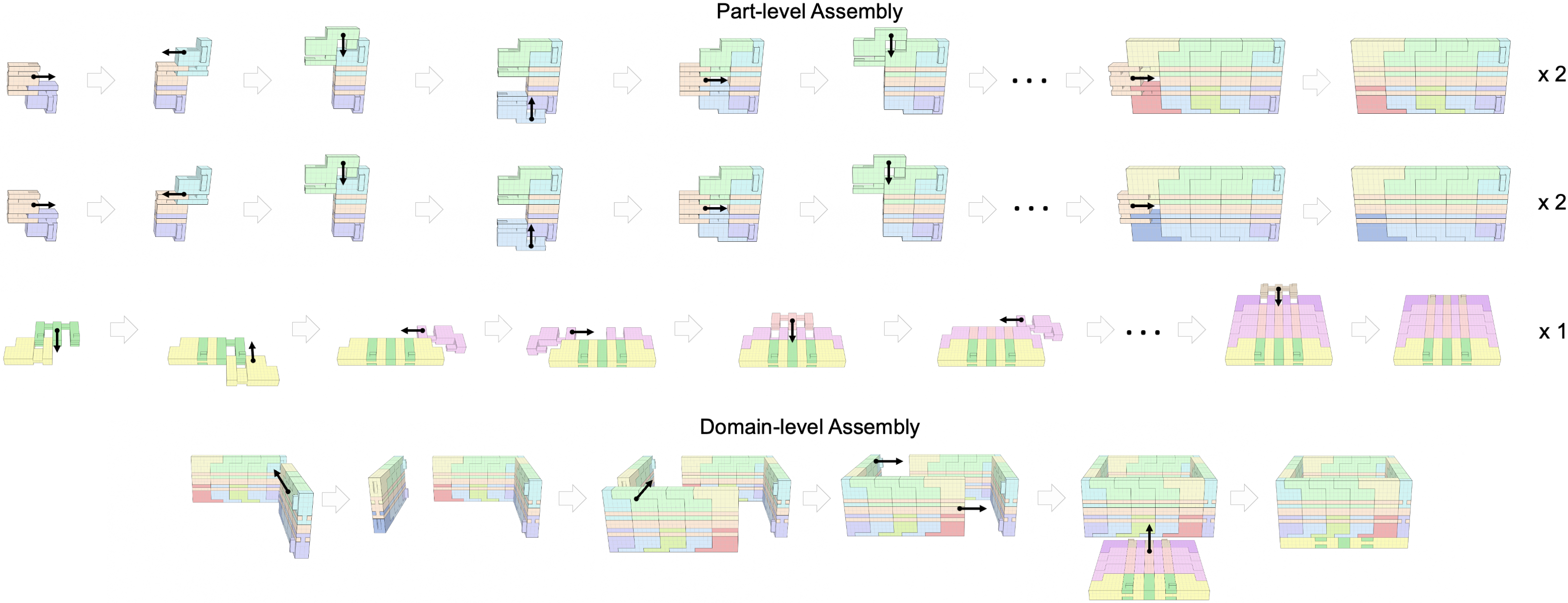}
	\caption{We show the two-level assembly process of an interlocking tiling \emph{Box} modeled by our approach: part-level (top) and domain-level (bottom) assembly.
	}
	\label{fig:result_assembly}
\end{figure*}

\begin{figure*}[!t]
	\centering
	\includegraphics[width=18.0cm]{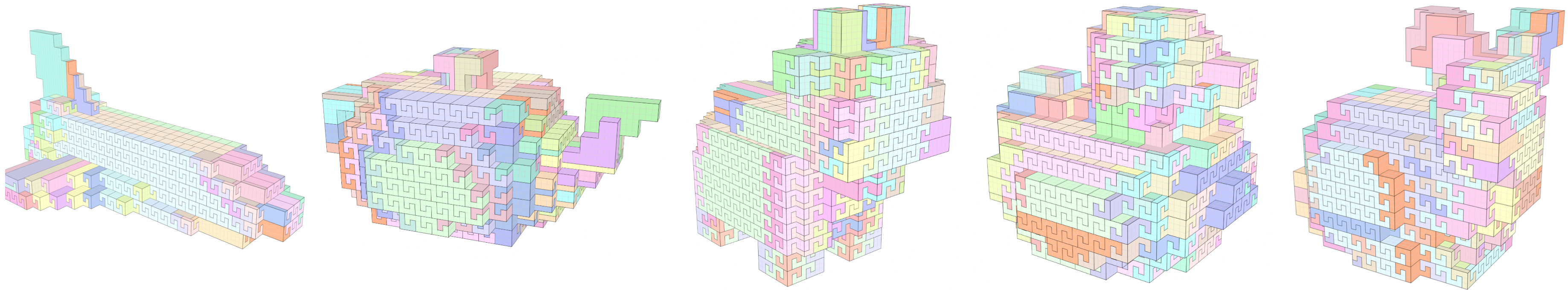}
	\caption{Our part-level partitioning approach is able to generate interlocking tiling results with complex boundaries: \emph{Shuttle}, \emph{Teapot}, \emph{Cow}, \emph{Duck}, and \emph{Bunny}.
	}
	\label{fig:result_solid_shape}
\end{figure*}

\begin{table}[!t]
	\centering
	\caption{
		Statistics and timings. We report
		the input domain ($D$),
		number of voxels in the domain ($M$),
		target number of tiles ($N_T^{\mathrm{target}}$),
		number of parts ($N_P$),
		number of tiles ($N_T$),
		number of prototiles in each interlocking tiling result ($K$),
		average reusability of computed prototiles ($N_T/K$),
		running time for domain-level partitioning ($T_\text{domain}$),
		part-level partitioning ($T_\text{part}$), and
		the entire interlocking tiling approach ($T_\text{total}$).
	}
	\includegraphics[width=8.35cm]{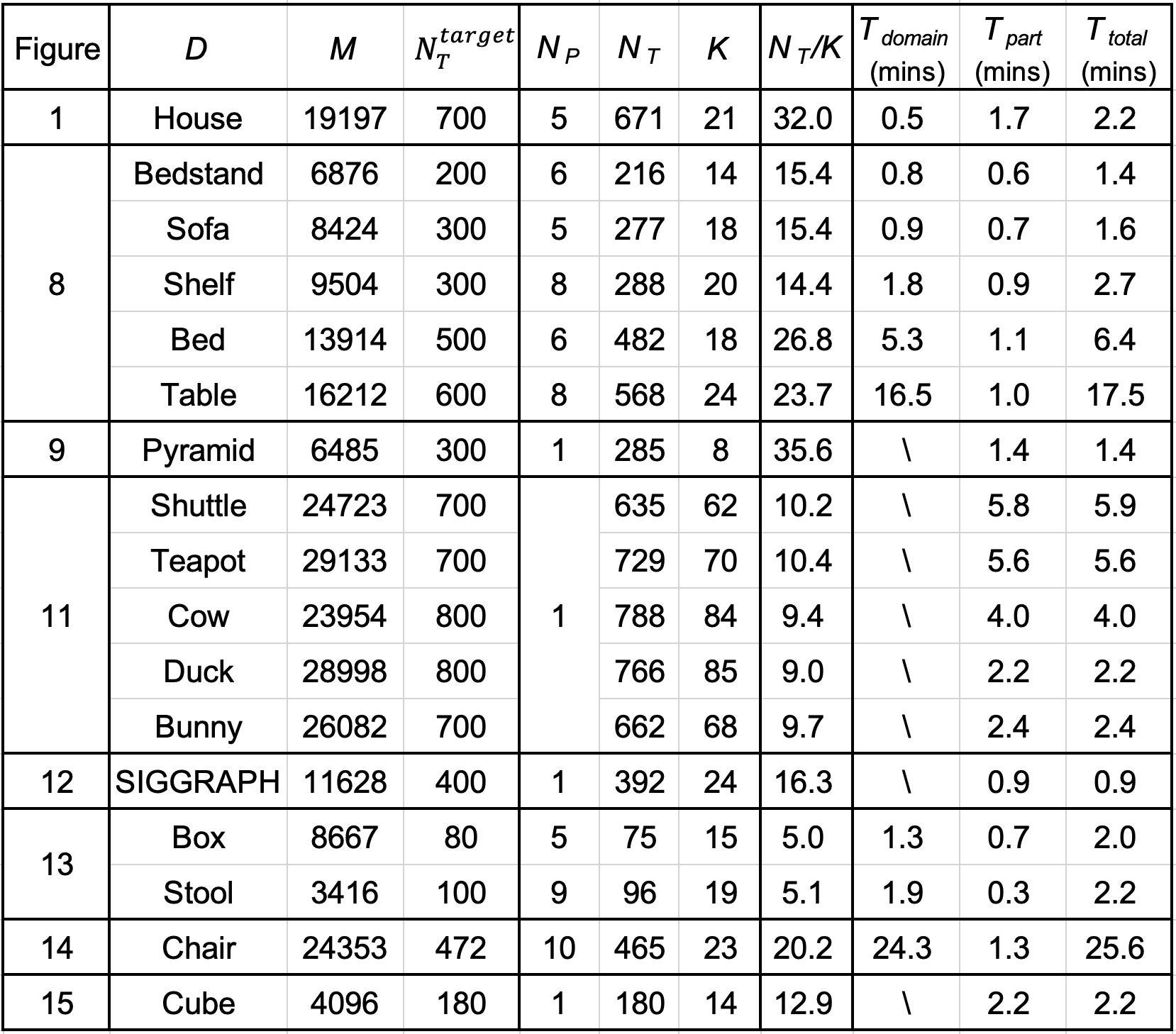}
	\label{tbl:result_statistics}
	\vspace*{-4.0mm}
\end{table}

We implemented our interlocking tiling approach in C++ and tested it on a workstation with a 4.06GHz CPU and 36GB memory.
To produce an interlocking tiling result, we need to specify the input domain $D$ and the target number of tiles $N_T^{\mathrm{target}}$.
For all our results, the resulting number of tiles $N_T$ deviates from $N_T^{\mathrm{target}}$ by less than $10\%$; see Table~\ref{tbl:result_statistics}.
To visualize the results, tiles with exactly the same shape are rendered in the same color.
Table~\ref{tbl:result_statistics} reports statistics for each result.
We also provide the 3D models of each interlocking tiling result as supplementary files.

\begin{figure*}[!t]
	\centering
	\includegraphics[width=18.0cm]{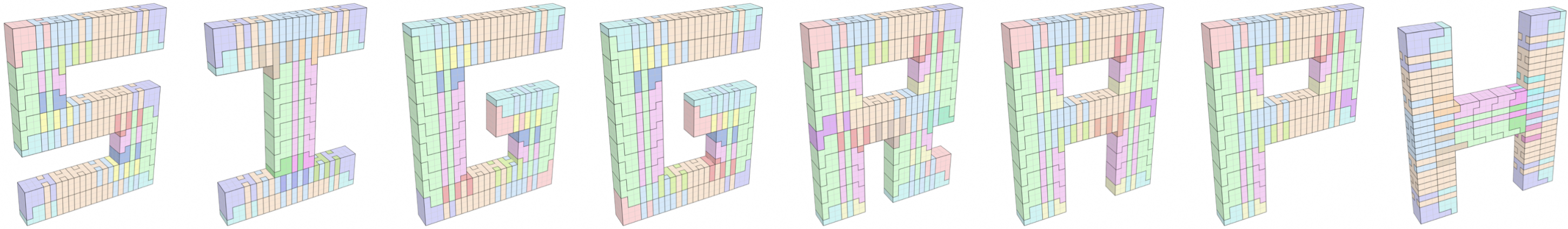}
	\caption{Our part-level partitioning approach is able to generate interlocking tiling results on a disconnected domain with eight components (i.e., eight letters).
	}
	\label{fig:result_disconnected}
\end{figure*}

\emph{Virtual results.}
We evaluated our interlocking tiling approach on a variety of voxelized input domains.
For each virtual result, the prototile reusability $N_T/K$ is consistently above 9.
Specifically, our results are categorized by the partitioning strategy employed:

	\begin{itemize}[leftmargin=*]
		\item
		{\em Two-level partitioning.} \
		Our two-level partitioning strategy can be used to produce interlocking tiling results on spatial domains; see Figure~\ref{fig:result_shape}.
		In these results, tiles are first assembled into interlocking parts, followed by assembling the parts into complete interlocking tiling results.
		Specifically, Figure~\ref{fig:result_assembly} shows a two-level assembly process of \emph{Box}.
		Please refer to the accompanying video for the assembly process of \emph{House} and \emph{Bedstand} produced by this strategy.

		\vspace*{1mm}
		\item
		{\em Part-level partitioning.} \
		If the domain is treated as a single part, the part-level partitioning approach can also produce interlocking tiling results for domains with complex boundaries and disconnected geometries; see Figures~\ref{fig:result_solid_shape} and~\ref{fig:result_disconnected}.
		These results highlight the flexibility of our part-level partitioning approach in adapting to geometrically complex domains through concatenation of multiple sets of IT-DBGs.
		Please refer to the accompanying video for the assembly process of \emph{Bunny} and \emph{Teapot} produced by this strategy.
	\end{itemize}

\begin{figure}[!t]
	\centering
	\includegraphics[width=0.85\linewidth]{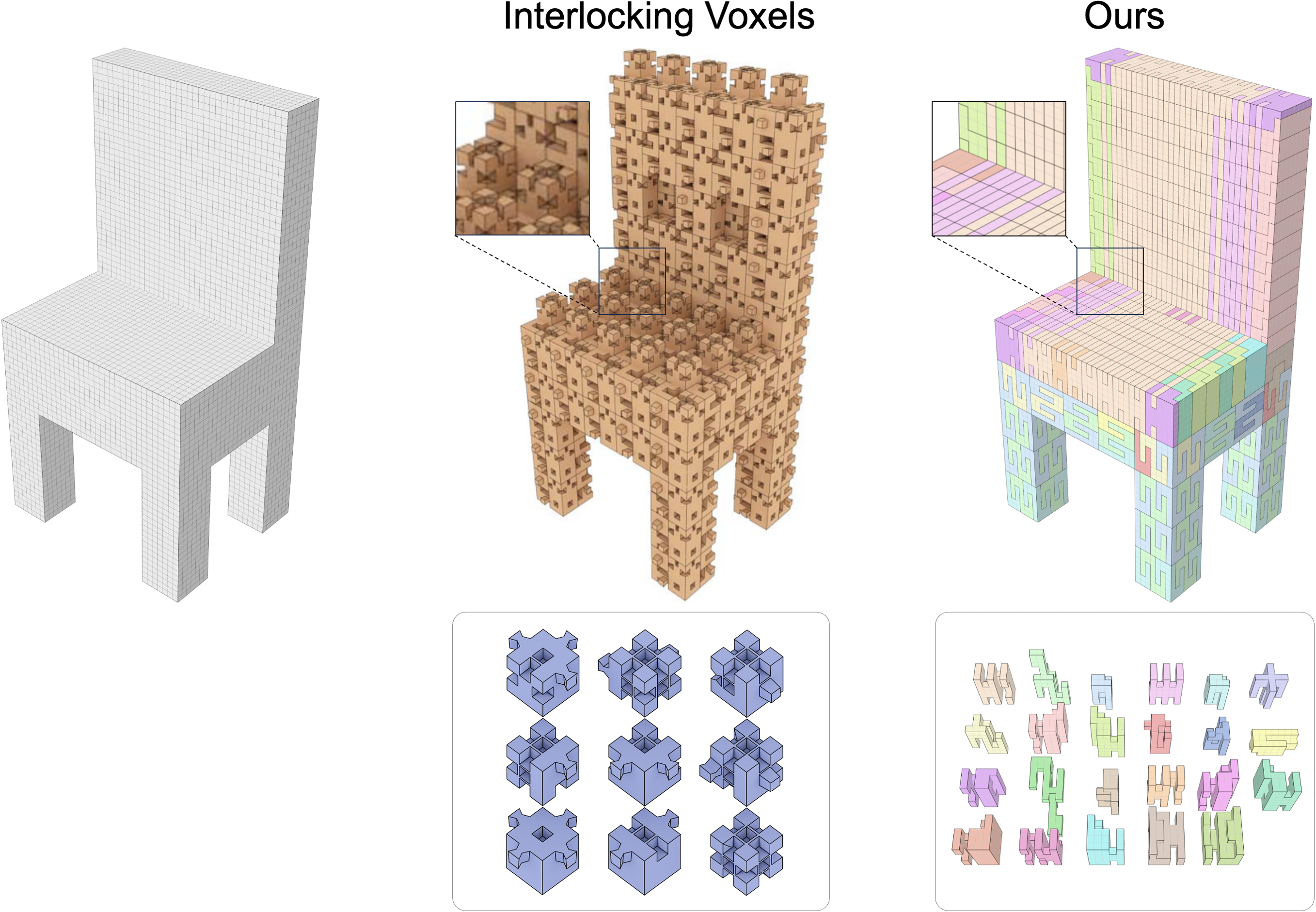}
	\caption{Comparing our approach with a forward modeling approach using interlocking voxels~\cite{Zhang-2021-InterlockBlock}.}
	\label{fig:result_compare_interlocking_voxel}
\end{figure}

\emph{Comparison with Interlocking Voxels~\cite{Zhang-2021-InterlockBlock}.}
We compare our interlocking tiling approach with a forward modeling approach using interlocking voxels~\cite{Zhang-2021-InterlockBlock}; see Figure~\ref{fig:result_compare_interlocking_voxel}.
Given the same input domain, their approach can only approximate the target shape, resulting in an assembly that contains many small cavities across the whole shape.
In contrast, although our computed interlocking tiling result has more unique tile shapes (23 vs. 9), it exactly covers the input domain, resulting in a nicer overall appearance.

\begin{figure}[!t]
	\centering
	\includegraphics[width=0.99\linewidth]{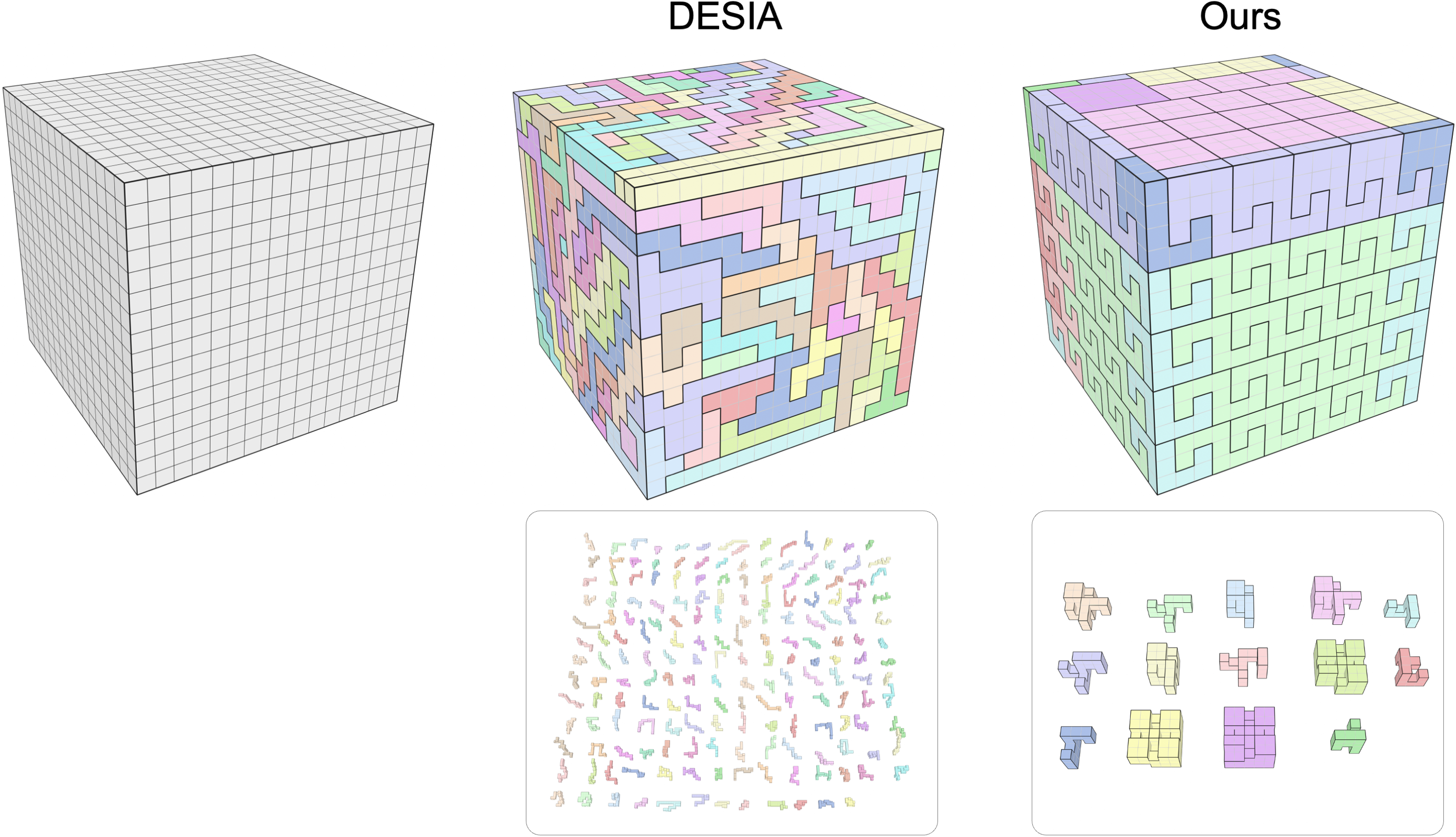}
	\caption{Comparing our approach with an inverse modeling approach called DESIA~\cite{Wang-2018-DESIA}.}
	\label{fig:result_compare_DESIA}
\end{figure}

\emph{Comparison with DESIA~\cite{Wang-2018-DESIA}.}
We compare our interlocking tiling approach with a DBG-guided modeling approach called DESIA~\cite{Wang-2018-DESIA}; see Figure~\ref{fig:result_compare_DESIA}.
Given the same input domain, DESIA generates an interlocking assembly that exactly covers the input domain, yet the resulting 472 tiles are all distinct.
In contrast, our approach produces an interlocking tiling result with only 14 unique tile shapes.

\begin{figure*}[!t]
	\centering
	\includegraphics[width=18.0cm]{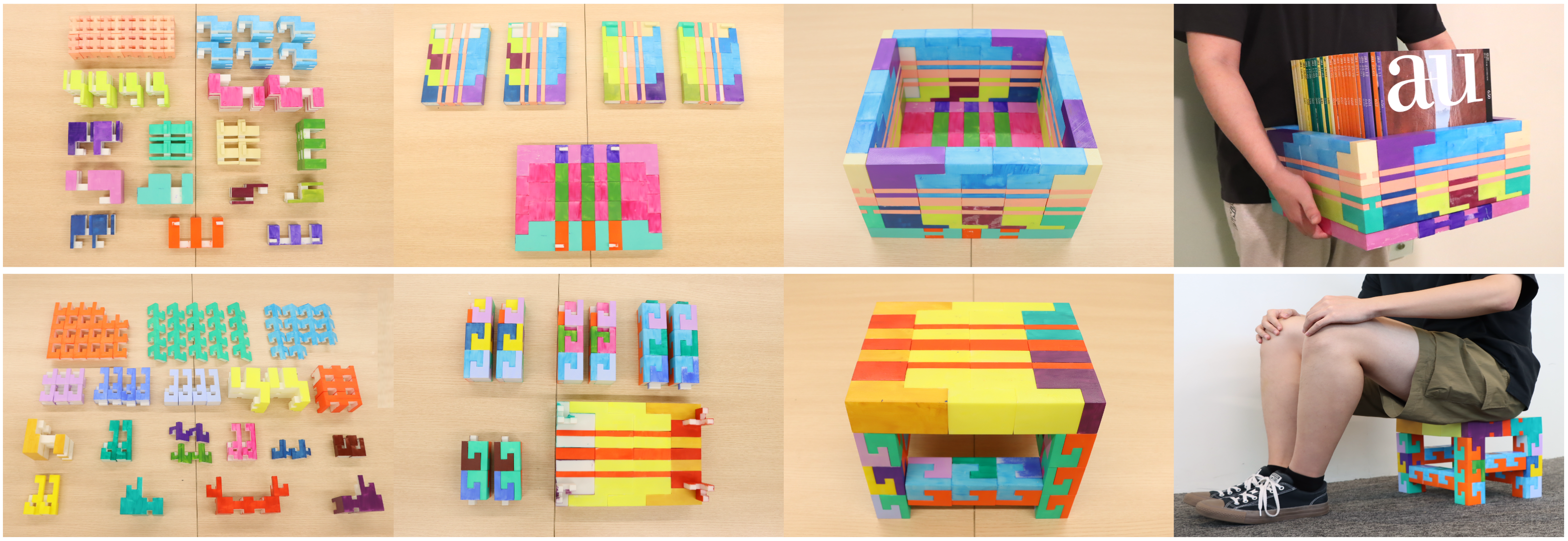}
	\caption{We fabricate interlocking tiling \emph{Box} (top) and \emph{Stool} (bottom) using 3D printing.
		The tiles with the same shape are painted using the same color.
		The physical \emph{Box} securely holds 15\,kg of books when lifted, and the physical \emph{Stool} safely supports a 90\,kg participant.
	}
	\label{fig:result_fabrication}
\end{figure*}

\emph{Physical results.}
We fabricated two of our interlocking tiling results using 3D printing and painted tiles with identical shapes using the same color; see Figure~\ref{fig:result_fabrication}.
We evaluate structural stability of these results through physical tests in real-world use scenarios.
The physical \emph{Box} (33\,cm $\times$ 33\,cm $\times$ 18\,cm) can be used to carry books with a total weight of approximately 15\,kg.
The physical \emph{Stool} (30\,cm $\times$ 25\,cm $\times$ 23.5\,cm) supports a user weighing approximately 90\,kg.
Please refer to the accompanying video for the physical assembly process and real-world usage of the \emph{Box} and \emph{Stool}.

\section{Conclusion}
\label{sec:conclusion}

This paper introduces \emph{interlocking tiling} of 3D finite domains, offering a new perspective on modeling physically-realizable 3D $K$-hedral tilings by jointly minimizing the number of prototiles and ensuring structural stability.
We propose a two-level hierarchical approach that decomposes the domain into interlocking parts and further decomposes each part into interlocking tiles.
In particular, we propose \emph{Interlocking Tiling Directional Blocking Graphs} to guide the generation of interlocking tiles with a small number of unique shapes.
Experimental results demonstrate that our approach can tile a wide variety of 3D finite domains and achieve superior performance compared with two state-of-the-art approaches.
We also fabricated two of our interlocking tiling results to validate their structural stability and usability in the real world.

\emph{Limitation and future work.}
First, our approach assumes that the input domain is represented as a voxel grid.
An interesting direction for future work is to generalize our approach to support domains with other representations, such as tetrahedral and hexahedral discretizations.
Second, our interlocking tiling approach requires the input shape to have a sufficiently large internal volume.
Otherwise, it may be difficult to ensure interlocking and/or to achieve high prototile reuse.
Third, we currently do not analyze the structural implications of the connections between individual tiles.
It would be beneficial to optimize the stress distribution within the assembly to avoid local stress concentrations, thereby extending our approach toward real-world, large-scale applications.

\bibliographystyle{IEEEtran}
\bibliography{Ref_InterlockTiling}

@STRING{TOG_SIG = {ACM Trans. on Graph. (SIGGRAPH)}}

@STRING{TOG_SIGA = {ACM Trans. on Graph. (SIGGRAPH Asia)}}

@STRING{SIGGRAPH = {Proc. of SIGGRAPH}}

@STRING{TVCG = {IEEE Trans. Vis. \& Comp. Graphics}}

@STRING{CGF_EG = {Comp. Graph. Forum (Eurographics)}}

@STRING{CGF_PG = {Comp. Graph. Forum (Pacific Graphics)}}

@STRING{CAG_SMI = {Comp. \& Graph. (SMI)}}

@STRING{CAGD_GMP = {Comp. Aided Geom. Des. (GMP)}}

@STRING{IROS = {Proc. of IEEE/RSJ Int. Conf. on Intelligent Robots and Systems}}

@STRING{UIST = {Proc. of ACM UIST}}

@inproceedings{Song-2022-InterlockingSurvey,
	author       =     {Peng Song},
	title           =    {Interlocking Assemblies: Applications and Methods},
	booktitle    =    {Materials Today: Proceedings (International Conference on Additive Manufacturing for a Better World)},
	volume     =    {70},
	year         =    {2022},
	pages      =    {78--82},
}

@inproceedings{Sun-2024-StructCurves,
	author       =     {Zezhou Sun and Devin Balkcom and Emily Whiting},
	title            =    {{StructCurves}: Interlocking Block-Based Line Structures},
	booktitle    =    UIST,
	year          =    {2024},
	pages       =    {39:1--39:11},
}

@inproceedings{Sniffen-2021-FallingIntoPlace,
	author       =     {Amy Sniffen and Zezhou Sun and Samuel Lensgraf and Emily Whiting and Alberto Quattrini Li and Devin Balkcom},
	title           =     {{Falling Into Place}: Drop Assembly of Interlocking Puzzles},
	booktitle   =     {Proc. of Robotics: Science and Systems},
	year         =     {2021},
}

@article{Zhang-2021-InterlockBlock,
	author        =    {Yinan Zhang and Yotto Koga and Devin Balkcom},
	title            =    {Interlocking Block Assembly With Robots},
	journal       =   {IEEE Transactions on Automation Science and Engineering},
	year          =    {2021},
	volume      =    {18},
	number     =    {3},
	pages       =    {902--916},
}

@inproceedings{Zhang-2016-InterlockVoxel,
	title            =    {Interlocking Structure Assembly with Voxels},
	author        =    {Yinan Zhang and Devin Balkcom},
	booktitle     =    IROS,
	pages        =    {2173--2180},
	year           =    {2016},
}

@inproceedings{Shih-2016-SLBlocks,
	author       =    {Shen-Guan Shih},
	title           =    {On the Hierarchical Construction of {SL} Blocks},
	booktitle    =    {Proc. of Advances in Architectural Geometry},
	pages        =    {124--136},
	year          =    {2016},
}

@inproceedings{Shih-2018-SLBlocks,
	author      =    {Shen-Guan Shih},
	title          =    {The Art and Mathematics of Self-Interlocking {SL} Blocks},
	booktitle   =    {Proc. of Bridges: Mathematics, Art, Music, Architecture, Education, Culture},
	pages       =    {107--114},
	year         =     {2018},
}

@inproceedings{Shih-2020-SLBlockGrammars,
	author      =    {Shen-Guan Shih},
	title          =    {Grammars of Interlocking {SL} Blocks},
	booktitle  =    {Proc. of Advances in Architectural Geometry},
	pages      =    {506--520},
	year        =     {2020},
}

@article{Chen-2022-HighLevelPuzzle,
	author     =    {Rulin Chen and Ziqi Wang and Peng Song and Bernd Bickel},
	title         =    {Computational Design of High-level Interlocking Puzzles},
	journal    =    TOG_SIG,
	volume   =    {41},
	number   =    {4},
	pages     =    {150:1--150:15},
	year       =    {2022}
}

@article{Tang-2019-3DDissection,
	author    =    {Keke Tang and Peng Song and Xiaofei Wang and Bailin Deng and Chi-Wing Fu and Ligang Liu},
	title         =    {Computational Design of Steady {3D} Dissection Puzzles},
	journal    =    CGF_EG,
	year        =    {2019},
	volume   =    {38},
	number  =    {2},
	pages     =    {291--303},
}

@article{Wang-2018-DESIA,
	author     =    {Ziqi Wang and Peng Song and Mark Pauly},
	title          =    {{DESIA}: A General Framework for Designing Interlocking Assemblies},
	journal    =    TOG_SIGA,
	year         =    {2018},
	volume   =    {37},
	number   =    {6},
	pages      =    {191:1--191:14},
}

@article{Song-2017-ReconfigInterlock,
	author     =    {Peng Song and Chi-Wing Fu and Yueming Jin and Hongfei Xu and Ligang Liu and Pheng-Ann Heng and Daniel Cohen-Or},
	title          =    {Reconfigurable Interlocking Furniture},
	journal    =    TOG_SIGA,
	volume    =    {36},
	number   =    {6},
	year         =    {2017},
	pages     =    {174:1--174:14},
}

@article{Yao-2017-InterlockShell,
	author      =    {Miaojun Yao and Zhili Chen and Weiwei Xu and Huamin Wang},
	title           =    {Modeling, Evaluation and Optimization of Interlocking Shell Pieces},
	journal     =    CGF_PG,
	volume    =    {36},
	number   =    {7},
	year         =    {2017},
	pages      =    {1--13},
}

@article{Fu-2015-Furniture,
	author      =    {Chi-Wing Fu and Peng Song and Xiaoqi Yan and Lee Wei Yang and Pradeep Kumar Jayaraman and Daniel Cohen-Or},
	title           =    {Computational Interlocking Furniture Assembly},
	journal      =    TOG_SIG,
	volume     =    {34},
	number    =    {4},
	year          =    {2015},
	pages       =    {91:1--91:11},
}

@article{Song-2015-Interlock,
	title           =    {Printing {3D} Objects with Interlocking Parts},
	author     =    {Peng Song and Zhongqi Fu and Ligang Liu and Chi-Wing Fu},
	journal     =    CAGD_GMP,
	volume    =    {35-36},
	pages      =    {137--148},
	year          =    {2015},
}

@article{Song-2012-InterCubes,
	author     =    {Peng Song and Chi-Wing Fu and Daniel Cohen-Or},
	title          =    {Recursive Interlocking Puzzles},
	journal    =    TOG_SIGA,
	year        =    {2012},
	volume   =    {31},
	number  =    {6},
	pages      =    {128:1--128:10},
}

@article{Xin-2011-BurrPuzzles,
	author      =    {Shi-Qing Xin and Chi-Fu Lai and Chi-Wing Fu and Tien-Tsin Wong and Ying He and Daniel Cohen-Or},
	title           =    {Making Burr Puzzles from {3D} Models},
	journal     =    TOG_SIG,
	year          =    {2011},
	volume    =    {30},
	number   =    {4},
	pages      =    {97:1--97:8},
}

@incollection{Dyskin-2019-TIMaterials,
	author        =     {A. V. Dyskin and Yuri Estrin and E. Pasternak},
	title             =    {Topological Interlocking Materials},
	editor         =     {Yuri Estrin and Yves Br{\'e}chet and John Dunlop and Peter Fratzl},
	booktitle    =      {Architectured Materials in Nature and Engineering},
	publisher   =      {Springer International Publishing},
	year           =     {2019},
	pages        =     {23--49},
	chapter     =     {2},
}

@article{Weizmann-2017-BuildingFloors,
	author     =    {Michael Weizmann and Oded Amir and Yasha Jacob Grobman},
	title          =    {Topological Interlocking in Architecture: A New Design Method and Computational Tool for Designing Building Floors},
	journal     =    {International Journal of Architectural Computing},
	year         =    {2017},
	volume    =    {15},
	number   =    {2},
	pages      =    {107--118},
}

@article{Dyskin-2003-OsteomorphicBlock,
	author      =    {A. V. Dyskin and Y. Estrin and E. Pasternak and H. C. Khor and A. J. Kanel-Belov},
	title          =    {Fracture Resistant Structures Based on Topological Interlocking with Non-planar Contacts},
	journal     =    {Advanced Engineering Materials},
	year         =    {2003},
	volume    =    {5},
	number   =    {3},
	pages     =    {116--119},
}

@article{Dyskin-2003-TIPlatonicSolids,
	author     =    {A. V. Dyskin and Y. Estrin and A. J. Kanel-Belov and E. Pasternak},
	title         =    {Topological Interlocking of Platonic Solids: A Way to New Materials and Structures},
	journal    =    {Philosophical Magazine Letters},
	year        =    {2003},
	volume    =    {83},
	number   =    {3},
	pages     =    {197--203},
}

@article{Wang-2019-TopoLock,
	author     =    {Ziqi Wang and Peng Song and Florin Isvoranu and Mark Pauly},
	title         =    {Design and Structural Optimization of Topological Interlocking Assemblies},
	journal    =    TOG_SIGA,
	year        =    {2019},
	volume    =    {38},
	number   =    {6},
	pages      =    {193:1--193:13},
}

@article{Bejarano-2019-TIDesignFramework,
	author      =    {Andres Bejarano and Christoph Hoffmann},
	title           =    {A Generalized Framework for Designing Topological Interlocking Configurations},
	journal      =    {International Journal of Architectural Computing},
	year          =    {2019},
	volume     =    {17},
	number    =    {1},
	pages      =    {53--73},
}

@incollection{Fallacara-2019-TIForArchitecture,
	author        =     {Giuseppe Fallacara and Maurizio Barberio and Micaela Colella},
	title             =    {Topological Interlocking Blocks for Architecture: From Flat to Curved Morphologies},
	editor         =     {Yuri Estrin and Yves Br{\'e}chet and John Dunlop and Peter Fratzl},
	booktitle    =     {Architectured Materials in Nature and Engineering},
	publisher   =     {Springer International Publishing},
	year           =    {2019},
	pages        =    {423--445},
	chapter     =    {14},
}

@article{Krishnamurthy-2022-Space-FillingTiling,
	author      =    {Vinayak R. Krishnamurthy and Ergun Akleman and Sai Ganesh Subramanian and Matthew Ebert and Jiaqi Cui and Chia-an Fu and Courtney Starrett},
	title           =    {Geometrically Interlocking Space-Filling Tiling Based on Fabric Weaves},
	journal      =    TVCG,
	year          =    {2022},
	volume     =    {28},
	number    =    {10},
	pages      =    {3391--3404},
}

@inproceedings{Krishnamurthy-2020-Bi-AxialWovenTiles,
	author       =     {Vinayak R. Krishnamurthy and Ergun Akleman and Sai Ganesh Subramanian and Katherine Boyd and Chia-An Fu and Matthew Ebert and Courtney Starrett and Neeraj Yadav},
	title            =     {{Bi-Axial Woven Tiles}: Interlocking Space-Filling Shapes Based on Symmetries of Bi-Axial Weaving Patterns},
	booktitle     =    {Proc. of Graphics Interface},
	pages        =    {286--298},
	year           =     {2020},
}

@article{Akleman-2020-GeneralizedAbeilleTiles,
	author      =    {Ergun Akleman and Vinayak R. Krishnamurthy and Chia-An Fu and Sai Ganesh Subramanian and Matthew Ebert and Matthew Eng and Courtney Starrett and Haard Panchal},
	title           =    {{Generalized Abeille Tiles}: Topologically Interlocked Space-filling Shapes Generated Based on Fabric Symmetries},
	journal      =    CAG_SMI,
	year          =    {2020},
	volume     =    {89},
	pages       =    {156--166},
}

@article{Subramanian-2019-DelaunayLofts,
	author      =    {Sai Ganesh Subramanian and Mathew Eng and Vinayak R. Krishnamurthy and Ergun Akleman},
	title           =    {{Delaunay Lofts}: A Biologically Inspired Approach for Modeling Space Filling Modular Structures},
	journal      =    CAG_SMI,
	year          =    {2019},
	volume     =    {82},
	pages       =    {73--83},
}

@book{Kaplan-2009-TilingTheory,
	author          =       {Craig S. Kaplan},
	title              =       {Introductory Tiling Theory for Computer Graphics},
	year             =       {2009},
	publisher     =       {Springer},
}

@inproceedings{Chen-2025-InverseTiling,
	author       =     {Rulin Chen and Xuyang Ma and Praveer Tewari and Chi-Wing Fu and Peng Song},
	title            =     {Inverse Tiling of 2D Finite Domains},
	booktitle     =    {Proc. of SIGGRAPH Asia},
	year           =     {2025},
	pages  = {43:1--43:11},
}

@article{Garvie-2022-TilingFiniteRegions,
	author        =      {Marcus R. Garvie and John Burkardt},
	title            =     {A Parallelizable Integer Linear Programming Approach for Tiling Finite Regions of the Plane with Polyominoes},
	journal       =     {Algorithms},
	volume      =      {15},
	number     =      {5},
	year          =     {2022},
	pages       =      {164:1--164:41},
}

@article{Garvie-2020-TilingFiniteRegions,
	author        =      {Marcus R. Garvie and John Burkardt},
	title            =     {A New Mathematical Model for Tiling Finite Regions of The Plane with Polyominoes},
	journal       =     {Contributions to Discrete Mathematics},
	volume      =      {15},
	number     =      {2},
	year          =     {2020},
	pages       =      {95--131},
}

@phdthesis{Wilson-1992-GeometricAssemblyPlanning,
  author       = {Wilson, Randall H.},
  title        = {On Geometric Assembly Planning},
  school       = {Stanford University},
  year         = {1992},
  type         = {Ph.D. Dissertation},
}

\makeatletter
\raggedbottom
\renewcommand{\@textbottom}{\vskip \z@ \@plus 1fill}
\makeatother

\begin{IEEEbiography}
	[{\includegraphics[width=1in,height=1.25in,clip,keepaspectratio]{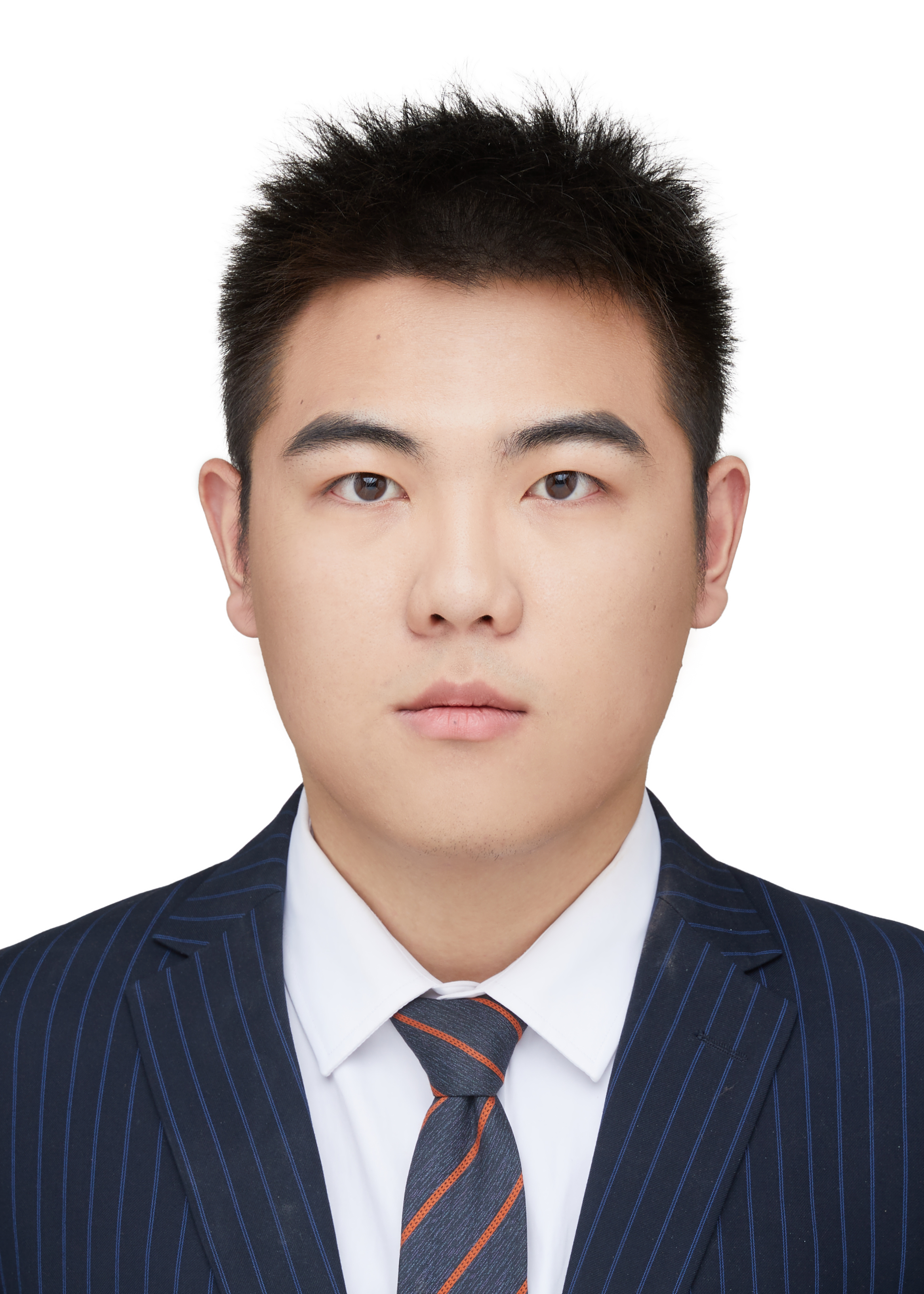}}]{Xuyang Ma}
	received the B.S. degree in software engineering from Shandong University, China, in 2021, and the M.S. degree from the Institute of Computing Technology, Chinese Academy of Sciences, China.
	He is currently a Ph.D. student with the Singapore University of Technology and Design, Singapore.
	His research focuses on the intersection of AI and computational design.
\end{IEEEbiography}

\begin{IEEEbiography}
	[{\includegraphics[width=1in,height=1.25in,clip,keepaspectratio]{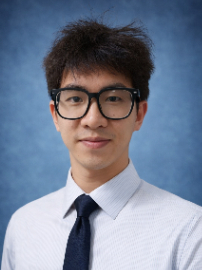}}]{Rulin Chen}
	received the bachelor's degree from Shantou University, in 2020 and the doctoral degree in 2024 under the supervision of Prof. Peng Song.
	He is an Assistant Professor in the School of Culture and Creativity, Beijing Normal-Hong Kong Baptist University.
	His research interest is in computer graphics, with a focus on computational assemblies. He received SIGGRAPH Technical Papers Award Honorable Mention, in 2022.
\end{IEEEbiography}

\begin{IEEEbiography}
	[{\includegraphics[width=1in,height=1.25in,clip,keepaspectratio]{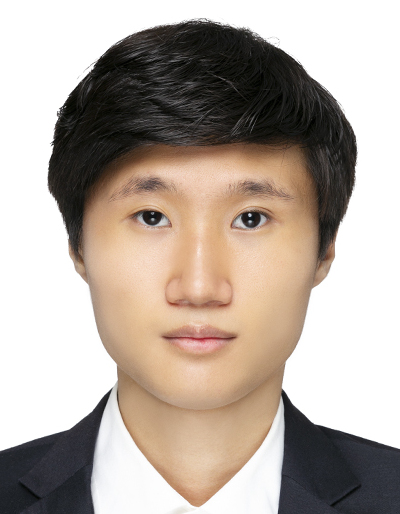}}]{Chaewoon Kim}
	received the Bachelor of Science degree in Computer Science and Design from the Singapore University of Technology and Design (SUTD), Singapore, in 2024. He is currently pursuing the PhD degree in Computer Science and Design at SUTD under the supervision of Prof. Peng Song. He is a member of the Computational Design Lab at SUTD. His research interests include computer graphics and computational design.
\end{IEEEbiography}

\begin{IEEEbiography}
	[{\includegraphics[width=1in,height=1.25in,clip,keepaspectratio]{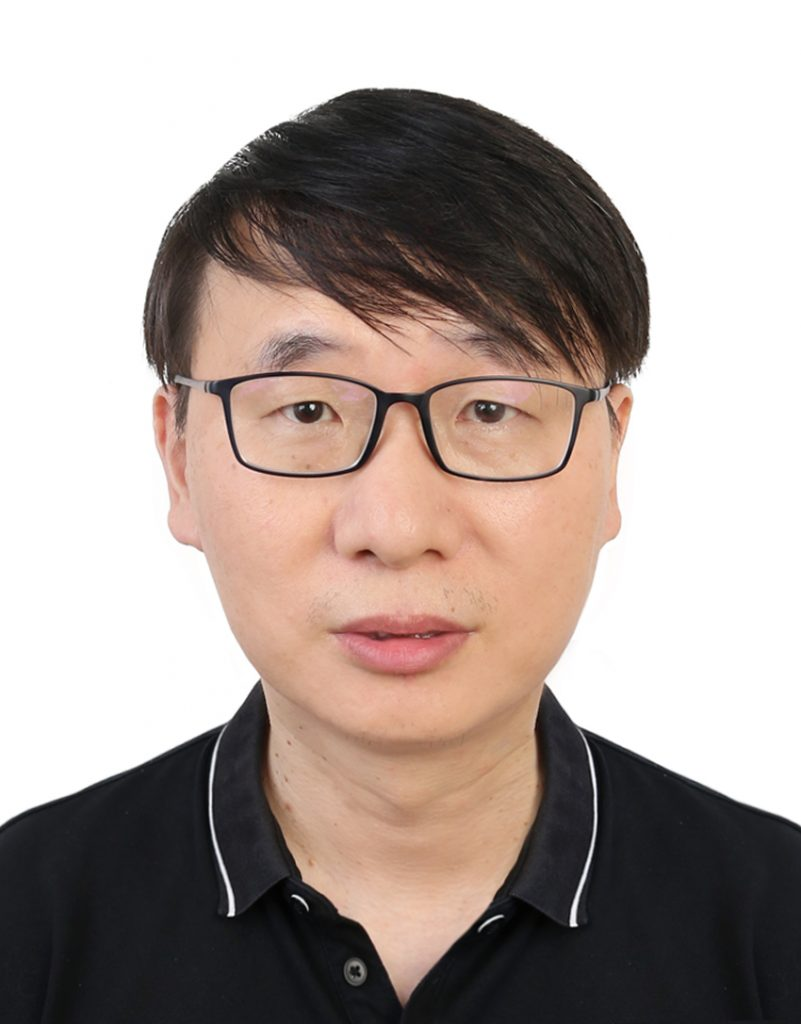}}]{Ying He} (Member, IEEE)
	received the bachelor's and master's degrees in electrical engineering from Tsinghua University, and PhD degree in computer science from Stony Brook University. He is an Associate Professor in the School of Computer Science and Engineering, Nanyang Technological University, Singapore. His research interests fall into the general areas of visual computing and he is particularly interested in the problems which require geometric analysis and computation. He is an associate editor of Journal of Computational Visual Media (Springer) and Computer Graphics Forum (Wiley). He regularly serves on the Technical Program Committee for major research conferences in geometric modeling and computer graphics.
\end{IEEEbiography}

\begin{IEEEbiography}
	[{\includegraphics[width=1in,height=1.25in,clip,keepaspectratio]{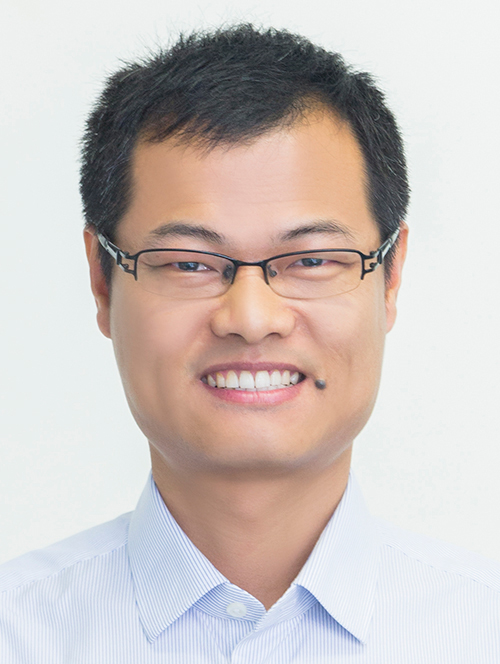}}]{Peng Song}
	received bachelor's and master's degrees from the Harbin Institute of Technology, China, in 2007 and 2010, respectively, and the PhD degree from Nanyang Technological University, Singapore, in 2013. He is an Assistant Professor with the Pillar of Information Systems Technology and Design, SUTD since 2019. Prior to joining SUTD, Peng was a research scientist with EPFL, Switzerland. His research interest is in geometric computing, with a focus on computational design and physical AI. He is an associate editor of Computers \& Graphics (Elsevier) and Graphical Models (Elsevier). He served as Posters Chair of SIGGRAPH Asia 2026.
\end{IEEEbiography}

\end{document}